\documentclass[%
 reprint,
  amsmath,amssymb,
 prb,
 ]{revtex4-1}

\usepackage{amsfonts,amsmath,latexsym,amssymb}
\usepackage{graphicx}% Include figure files
\usepackage{color}
\usepackage{dcolumn}% Align table columns on decimal point
\usepackage{bm}% bold math
\usepackage{hyperref}% add hypertext capabilities
\usepackage{enumerate}

\definecolor{bronze}{rgb}{0.8, 0.5, 0.2}

\begin{document}

%\preprint{APS/123-QED}

%%%%%%%%%%%%%%%%%%%%%%%%%%%%%%%%%%%%%%%%%%%%%%%%%%%%%%%%%%%%%%%%%%%%%%%%%%%%%%%%%%%%%%%%%%%%%%%%%%%%%%%%%%%%
\title{Charge and spin correlations in the Monopole Liquid}
%%%%%%%%%%%%%%%%%%%%%%%%%%%%%%%%%%%%%%%%%%%%%%%%%%%%%%%%%%%%%%%%%%%%%%%%%%%%%%%%%%%%%%%%%%%%%%%%%%%%%%%%%%%%

   \author{D. Slobinsky}
   \affiliation{Instituto de F\'{\i}sica de L\'{\i}quidos y Sistemas Biol\'ogicos (IFLYSIB), UNLP-CONICET, La Plata, Argentina.}
   \affiliation{Departamento de Ingenier\'ia Mec\'anica, Facultad Regional La Plata, Universidad Tecnol\'ogica Nacional, Av. 60 Esq. 124, 1900 La Plata, Argentina}% 
 %\email{Second.Author@institution.edu}
   
   \author{G. Baglietto}
   \affiliation{Instituto de F\'{\i}sica de L\'{\i}quidos y Sistemas Biol\'ogicos (IFLYSIB), UNLP-CONICET, La Plata, Argentina}

   \author{R. A. Borzi}
   \affiliation{Instituto de F\'{\i}sica de L\'{\i}quidos y Sistemas Biol\'ogicos (IFLYSIB), UNLP-CONICET, La Plata, Argentina}
   \affiliation{Departamento de F\'\i{}sica, Facultad de Ciencias Exactas, Universidad 
   Nacional de La Plata,  c.c.\ 16, suc.\ 4, 1900 La Plata, Argentina.}
%\ead{{borzi@fisica.unlp.edu.ar}}

\date{\today}

%%%%%%%%%%%%%%%%%%%%%%%%%%%%%%%%%%%%%%%%%%%%%%%%%%%%%%%%%%%%%%%%%%%%%
%%%%%%%%%%%%%%%%%%%%%%%%%   Abstract   %%%%%%%%%%%%%%%%%%%%%%%%%%%%%%
%%%%%%%%%%%%%%%%%%%%%%%%%%%%%%%%%%%%%%%%%%%%%%%%%%%%%%%%%%%%%%%%%%%%%
%%%%%%%%%%%%%%%%%%%%%%%%%%%%%%%%%%%%%%%%%%%%%%%%%%%%%%%%%%%%%%%%%%%%%
\begin{abstract}
A \emph{monopole liquid} is a spin system with a high density of magnetic charges, but no magnetic-charge order. We study such a liquid over an Ising pyrochlore lattice, where a single topological charge or \emph{monopole} sits in each tetrahedron. Restricting the study to the case with no magnetic field applied we show that, in spite of the liquid-like correlations between charges imposed by construction constraints, the spins are uncorrelated like in a perfect paramagnet. We calculate a massive residual entropy for this phase ($\ln(2)/2$, a result which is exact in the thermodynamic limit), implying a free Ising-like variable per tetrahedron. After defining a simple model Hamiltonian for this system (the \emph{balanced} monopole liquid) we study its thermodynamics. Surprisingly, this monopole liquid remains a perfect paramagnet at all temperatures. Thermal disorder can then be simply and quantitatively interpreted as single charge dilution, by the excitation of neutral sites and double monopoles. The addition of the usual nearest neighbors interactions favoring neutral `2in-2out' excitations as a perturbation maintains the same ground state, but induces short-range (topological) order by thermal disorder. While it decreases charge-charge correlations, pair spin correlations ---resembling those in spin ice--- appear on \emph{increasing temperature}. This helps us to see in another light the dipolar-like correlations present in spin ices at unexpectedly high temperatures. On the other side, favoring double excitations strengthen the charges short range order and its associated spin correlations. Finally, we discuss how the monopole liquid can be related to other systems and materials where different phases of \emph{monopole matter} have been observed.
\end{abstract}

%%Rule of thumb (write it or not?) 
% *charge correlations tend to be 
%  -created by double charges
%  -somewhat preserved by single charges
%  -erased by neutral sites
% *dipolar spin correlations tend to be 
%  -erased by double charges
%  -somewhat preserved by single charges
%  -created by neutral sites

\pacs{75.40.Cx, 75.10.Hk, 75.40.Mg,02.70.Uu,75.50.-y}

\maketitle

\vspace{2pc}
\noindent{\it Keywords}: spin ice, monopoles, spin liquid.

%%%%%%%%%%%%%%%%%%%%%%%%%%%%%%%%%%%%%%%%%%%%%%%%%%%%%%%%%%%%%%%%%%%%%
%%%%%%%%%%%%%%%%%%%%%%%%   Motivation   %%%%%%%%%%%%%%%%%%%%%%%%%%%%%
%%%%%%%%%%%%%%%%%%%%%%%%%%%%%%%%%%%%%%%%%%%%%%%%%%%%%%%%%%%%%%%%%%%%%
%%%%%%%%%%%%%%%%%%%%%%%%%%%%%%%%%%%%%%%%%%%%%%%%%%%%%%%%%%%%%%%%%%%%%
\section{Introduction}

%{\color{green} Discutir y ponerse de acuerdo en el texto: a qué llamamos exactamente ``paramagneto perfecto"? a- a un sistema que es paramagnético a toda temperatura? b- a un sistema que a una T dada no muestra correlación alguna (i.e., spines completamente independientes)? c- como b), pero a toda T?}
% Betts, introduction to mK technology: a system in which interactions between m.m. are ignored.
%Magnetism, Volume 1 edited by Etienne Du Trémolet de Lacheisserie: (...)this relation satisfied in the special case of a perfect paramagnet, defined as a material which obeys the Curie law M = ^H/T

About ten years ago, new quasiparticles with properties similar to magnetic monopoles were proposed to exist in spin ices \cite{Castelnovo2008}. Spin ice monopoles are excitations of the (disordered but highly correlated) magnetic ground state of these geometrically frustrated materials ---the magnetic analogs of water ice \cite{Ramirez1999,Bramwell2001}. Monopoles have four possible topological charges: single and double charge, both of which are either positive or negative. Like elementary particles with spin, single charges have internal degrees of freedom (their magnetic moment) while the more energetically costly double charges have none. At low temperatures and no applied field, monopoles exist only in low densities \cite{zhou2011,zhou2012chemical}, forming a charged fluid similar to diluted ionic solutions ---which indeed describe spectacularly well the physics of spin ice materials at low temperatures \cite{Morris2009,jaubert2009nat}. 
However, and interesting as it is, the physics of monopoles does not exhaust itself in providing a simpler and conceptual description of spin ice dynamics and thermodynamics. Indeed, as vortices in superconductors \cite{Blatter1994}, magnetic monopoles can be thought as atomic building blocks giving rise to new, dense phases of `monopole matter' \cite{Castelnovo2008,Sazonov2012,Borzi2013,BBartlett2014,Guruciaga2014,jaubert2015p,Rau2016}. A complete study of one of these phases in zero magnetic field ---the \emph{monopole liquid} (ML), a dense fluid of correlated magnetic charges that is also a perfect paramagnet--- is the main purpose of this work.

Very recently, it was proposed a new type of crystallography based on monopoles \cite{jaubert2015p}. This may sound exaggerated: the physics of a classical system of regular (electric) charges forced to inhabit the sites of a rigid lattice is quite interesting, but only a limited number of charge phases could be expected \cite{Dickman1999,Dickman2000,Borzi2013}. What new ingredients could exist in a system of magnetic monopoles that may lead to new phenomena? The answer to this question is manifold, and will be developed along this paper: $i-$ the interpretation of dipolar interactions among all spins as a Coulomb potential between monopoles is an approximation; residual effects can lead to degeneracy breaking (for example, the ordering of the monopole vacuum  \cite{Melko2001,Borzi2013}); $ii-$ even with no Coulomb interactions (i.e. negligible dipolar forces) there are still strong spin correlations that can lead to entropic forces among charges \cite{Castelnovo2012}, to the condensation of a crystal of double charges \cite{Guruciaga2014}, or to order by disorder \cite{Guruciaga2016}; $iii-$ spin interactions cannot \emph{in general} be interpreted fully in terms of charges; their effect is to select certain spin configurations, with non-trivial charge counterparts. Quite interestingly, second and third neighbor interactions between spins can be tuned to produce an effective pure charge repulsion, but also attraction between like charges \cite{Rau2016,Udagawa2016}; $iv-$ a magnetic field can be used to promote or depress the density of one type of charge at the expense of others \cite{Castelnovo2008,Sazonov2012,Guruciaga2016}; $v-$ due to the charges internal degrees of freedom, crystal formation is not in general the end of the story: a new type of magnetism on top of the magnetic crystal can lead to peculiar phenomena, as massive ground state degeneracy \cite{Borzi2013,jaubert2015s} and magnetic moment fragmentation \cite{BBartlett2014,Fennell2012,Petit2012,Petit2016,Lefrancois2017} vi- magnetic charges sit in a lattice where other degrees of freedom ---aside from the magnetic ones--- are likely to be active \footnote{Though we will not analyze this further, this effect may be at the core of the Balance Monopole Liquid Hamiltonian discussed in subsequent sections \cite{jaubert2015p,jaubert2015s,Khomskii2012}.}. Their coupling with the spin system can produce new effective magnetic interactions \cite{Khomskii2012,jaubert2015p,Fennell2014,Borzi2016,Penc2004,Bergman2006,tchernyshyov2011lacroix}.

In spite of much effort \cite{Pomaranski2013,Henelius2016,Borzi2016}, the ground state into which the spin ice materials (Dy$_2$Ti$_2$O$_7$, Ho$_2$Ti$_2$O$_7$ and others) freeze remains unknown. Conversely, different single monopole correlated states \cite{Castelnovo2008,Sakakibara2003,Sazonov2012,Borzi2013,Udagawa2016,Petit2016,Lefrancois2017} can be thought as phases crystallizing from a disordered dense fluid of monopoles: the monopole liquid. It is a phase that has been partially approached before \cite{Borzi2013,BBartlett2014}, but whose general properties have not yet been studied. This is the starting point of our paper. We begin by describing the ground state of the ML phase, a manifold populated by all spin configurations leading to a single monopole fluid of maximum density. We show that in spite of its spin disorder, massive residual entropy, and no interactions between charges, it has a liquid-like charge-charge correlation function. We give afterwards the arguments that lead to a specific Hamiltonian (with four-spin interactions) that is compatible with this ground state. We named this model ---which has the further attractive of admitting an analytic solution \cite{barry1989}--- the \emph{balanced} monopole liquid (BML), to stress the fact that it does not exhibit pair spin correlations at \emph{any} temperature. Including nearest neighbor interactions as a perturbation preserves the ground state properties but induces short-range spin correlations on \emph{increasing temperature}. While the `balanced liquid' case was intrinsically tuned to be a perfect paramagnet at all temperatures, we show that the perturbed cases constitute different versions of short-range spin order induced by disorder. Correlations can be made to develop towards the `all-in/all-out' state by antiferromagnetic perturbations, but also  towards the topological order measured in spin ices \cite{Fennell2009,Fennell2012} if the ferrogmanetic character is chosen. This study, in turn, will provide a simple route to understand in a microscopic way why dipolar-like correlations are observed in spin ices at temperatures as high as ten times the coupling constant \cite{Fennell2009,Chang2010}. 
%\textcolor{red}{Indeed, our model Hamiltonian leads naturally to a second perfect paramagnet with a ground state composed of neutral sites ---which tend to propagate dipolar spatial correlations--- and double monopoles ---in the exact proportion to erase them.}
Finally, we include a discussion on the effect of other perturbations on the ML, and on its possible relation with other systems and materials.

%%%%%%%%%%%%%%%%%%%%%%%%%%%%%%%%%%%%%%%%%%%%%%%%%%%%%%%%%%%%%%%%%%%%%
%%%%%%%%%%%%%%%%%%%%%%%%%%%   Model   %%%%%%%%%%%%%%%%%%%%%%%%%%%%%%%
%%%%%%%%%%%%%%%%%%%%%%%%%%%%%%%%%%%%%%%%%%%%%%%%%%%%%%%%%%%%%%%%%%%%%
%%%%%%%%%%%%%%%%%%%%%%%%%%%%%%%%%%%%%%%%%%%%%%%%%%%%%%%%%%%%%%%%%%%%%
%\section{Construction of the Model}
\section{Properties of the Monopole Liquid and a Model Hamiltonian: the \emph{Balanced} Monopole Liquid}

\subsection{Magnetic Lattice, Monopoles, and Nearest Neighbors Model}
\label{NNmodel}
\begin{figure}
    \includegraphics[width=0.45\textwidth]{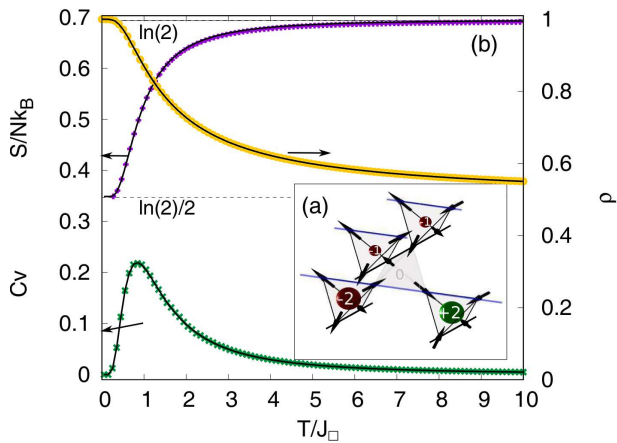}
    \caption{(a) A spin configuration in the pyrochlore lattice, composed of ``down" (center) and ``up" tetrahedra (the other four). Single and double charged monopoles (spheres), and neutral sites are labelled by their charges.
    (b) Monte Carlo simulations for the Balanced Monopole Liquid (BML) of Eq.~\ref{eq1} with $L = 8$ ($8192$ spins). In the right axis we show the charge density $\rho$ of single monopoles as a function of temperature obtained by simulations (yellow curve) and from the exact result \cite{barry1989} described in the text (black line). The low temperature value $\rho=1$ indicates one single monopole in each tetrahedron. In the left axis the calculated specific heat (green crosses) and, in perfect agreement, the exact result \cite{barry1989} (black line) show a broad Schottky anomaly near $T = J_{\square}$, at the onset of the monopole fluid state. The purple curve shows the entropy per spin, obtained by integration of $C_V/T$; the Pauling-like estimate $\ln(2)/2$ at $T=0$ is quite compatible with high temperature limit of $\ln(2)$. We have also plotted (black line) the entropy of mixing for single monopoles diluted with excitations ($\Delta S_{mix}(T)/Nk_B = -1/2 \times [\rho(T) \ln(\rho(T)) + (1-\rho(T)) \ln(1-\rho(T))]$, after adding the same residual entropy. The almost perfect coincidence between both entropies reflects that thermal disorder manifests only through the incertitude in the position of the excitations.}
    \label{fig1}
\end{figure}

Our base system, where the effective magnetic monopoles\footnote{Along this work we will refer to the analogous of magnetic monopoles both as `monopoles' or as `charges'.} live, consists of a pyrochlore lattice of classical Ising spins. These spins have $\langle 111 \rangle$ anisotropy, which implies that they can only point into or out of the center of the ``up" tetrahedra in which corners they sit (see Fig.~\ref{fig1}(a)). %It is known that spin ice basic properties are adequately described by the \emph{dipolar model} \cite{Melko2004}, including dipolar and first neighbors exchange interactions. 
If we consider only nearest neighbors interactions between spins, we have the simplest model describing the physics of spin ice systems (the \emph{nearest neighbors spin ice model}) given by:

\begin{equation}
    {\cal H}_{nn}/k_B=J_{nn}\sum_{<i,j>}\sigma_i\sigma_j \,,
\label{effnn}
\end{equation}

%\noindent where $J_{nn}=J/3+5D/3>0$ is the effective antiferromagnetic nearest neighbors coupling, $\sigma=\pm1$ is a
\noindent where $J_{nn} > 0$ is the effective nearest neighbors coupling,  $i,j$ sweep nearest neighbor lattice sites, $\sigma_i$ is a
pseudospin that takes the value $\sigma_i=1 (-1)$ if the spin point outwards (inwards) of its up tetrahedron. This Hamiltonian is an adequate starting point for our discussion on a ML model. A positive $J_{nn}$ favors ferromagnetic `in-out' pairing of magnetic moments \footnote{Although $J_{nn} > 0$ in the Ising model, we refer to this case as `ferromagnetic', since this is the case when magnetic moments (and not pseudospins) are considered \cite{moessner1998rapcomm}.} that translates into the \emph{ice rule}: two spins should point in and two out (2in-2out) of each tetrahedron ---a divergence-free-like rule, resembling a condition for charge neutrality \cite{Henley2010}. It conducts to a zero temperature residual entropy similar to that of water ice, with a highly correlated ground state \cite{Henley2010}. These correlations have been shown to exist in spin ice materials by the appeareance of anisotropic diffuse scattering and \emph{pinch points} in neutron experiments \cite{Fennell2007}.
Excitations from the 2in-2out manifold are unfulfillements of the `divergence free' ice rule, what in turn can be understood as creation of topological charge. The different ways of assembling a 3in-1out configuration constitute four types of positive \emph{single} charges, while the four 3out-1in are their negative versions. They sit on the center of tetrahedra, which build up the sites of a diamond lattice. The ice rule can be broken in a more extreme way (with a bigger divergence or topological charge): 4 spins in, or 4 out, are the two examples of \emph{double} magnetic excitations. The presence of charges affect the dipolar-like spin correlations characterizing the spin ice ground state.  However, their perturbation seems to be less effective than expected: temperatures ($T$) as big as $T \approx 10J_{nn}$ \cite{Fennell2009}, and impurities do not erase them completely \cite{Chang2010}. In first approximation, the adition of dipolar interactions to Eq.~\ref{effnn} does not change the degeneracy of the 2in-2out manifold. However, it activates an effective Coulomb potential between pairs of monopoles \cite{Castelnovo2008}.%, which at low temperatures dominates over a much weaker effective entropic interaction \cite{Castelnovo2012}. 

Changing the sign of $J_{nn}$ in Eq. \ref{effnn} results of course in an inversion of the energy cost of neutral (with zero divergence) sites, single, and double monopoles. Although no interactions between charges are implied by Eq. \ref{effnn}, the ground state is not a liquid, but a \textit{crystal} of double monopoles \cite{denHer2000}. The reason why it does crystallize has to do with spin constraints, that appear as charge correlations within the monopole picture \cite{Guruciaga2014}: the only way to construct a dense arrangements of 4-in and 4-out charges is to put a positive (negative) double charge next to a negative (positive) one, assembling the `all-in/all-out' (AIAO) spin configuration. This type of correlation is also at work for single charges \cite{Henley2010,Castelnovo2011}, leading to the possibility of order-by-disorder in these systems \cite{Guruciaga2014}.

\subsection{Model-independent properties of the Monopole Liquid}

Before proposing a model Hamiltonian with a ML ground state, it will be interesting to discuss first the peculiar properties of the ground state itself. In the thermodynamic limit, any ML liquid configuration can be defined by specifying the sign of one spin per tetrahedron (for example the sign of the $[111]$ pseudospin in each up tetrahedron, and that of every $[1\bar1\bar1]$ one ---that we temporarily assign to the corresponding down tetrahedra). 
Fig.~\ref{fig1}(a) shows these spins arranged along blue straight lines. Strong restrictions on the other pseudospins (sitting along perpendicular black lines in Fig.~\ref{fig1}(a)) are imposed by the connectivity of the pyrochlore lattice and the need to have one single monopole per tetrahedron. It is easy to see that, due to these constraints, we are only free to chose the sign of one pseudospin per black line. This contribution grows as $\exp(N^{2/3})$ and thus does not add up to the entropy per spin (or any thermodynamic observable) as $N \rightarrow \infty$. 
We have thus mapped the ML to a perfect spin-$1/2$ paramagnet with half of the spins. 
It is trivial to compute now the residual entropy per spin of the model which turns out to be $1/2\ln(2)$. This result coincides exactly with the estimate using Pauling's argument, calculated using eight allowed configurations per tetrahedron instead of the six present in spin ice \cite{Diep2004}.

One could naively expect the ground state to be a neutral gas of charges (given that there are no charge interactions). On the other hand, taking into account charge correlations ---which just by themselves are able to induce charge order for double monopoles \cite{Guruciaga2014}--- one can speculate on the condensation of a crystal of single monopoles \cite{BBartlett2014}; both cases would be compatible with a large residual entropy. We will see that in our case spin constraints are not restrictive enough to force a monopole crystal at low temperatures, but its high density and the expected charge positional correlations \cite{Castelnovo2012} makes the ML ground state better described as liquid-like than as a gas\footnote{Of course, quantum gases do show particle correlations; we opted anyway to call our classical system a liquid.}. To show this, we can fix the charge on a tetrahedron and count the number of spin configurations on the nearest neighbor tetrahedra; they appear on a ratio $10:6$ for neighbouring monopoles of the opposite and the same sign, respectively. It is easy to see that this implies that the average total charge around a positive monopole ($+1$) is $-1$.
%Indeed, counting spin configurations on neighboring tetrahedra with the charge of a central monopole fixed, \textcolor{red}{the amount of configurations having as nearest neighbour a monopole of the opposite sign or of the same one are in a ratio of 10:6}. 

Surprisingly, the spins underlying the monopoles appear to have much more freedom than these topological charges. Proceeding as before for charges, the correlation function for spins can be readily calculated by fixing an arbitrary pseudospin from the ML to be positive, and then asking for the probability that a nearest neighbor spin has the same sign. The answer to this question is $1/2$, implying that there is \emph{no correlation} between nearest neighbors and ---by connecting tetrahedra--- that the same is true for further neighbors. 
It will be instructive to compare these results for the ML with those for the spin ice ground state. There, fixing a spin direction biases that of a nearest neighbor: it is $1/3$ more likely to have a neighboring pseudospin with the opposite sign than with the same one \cite{Henley2010}. The result emanates directly from the divergence free condition: once a pseudospin is fixed in a tetrahedron the probability of having one with opposite sign becomes higher, so that the magnetic-like flux lines are not interrupted. This, in turn, gives rise to dipolar correlations \cite{Henley2010}.

%%%%%%%%%%%%%%%%%%%%%%%%%%%%%%%%%%%%%%%%%%%%%%%%%%%%%%%%%%%%%%%%%%%%%
%%%%%%%%%%%%%%%%%%%%%%    Neutrons     %%%%%%%%%%%%%%%%%%%%%%%%%%%%%%
%%%%%%%%%%%%%%%%%%%%%%%%%%%%%%%%%%%%%%%%%%%%%%%%%%%%%%%%%%%%%%%%%%%%%
%%%%%%%%%%%%%%%%%%%%%%%%%%%%%%%%%%%%%%%%%%%%%%%%%%%%%%%%%%%%%%%%%%%%%

\begin{figure}
    \includegraphics[width=0.45\textwidth]{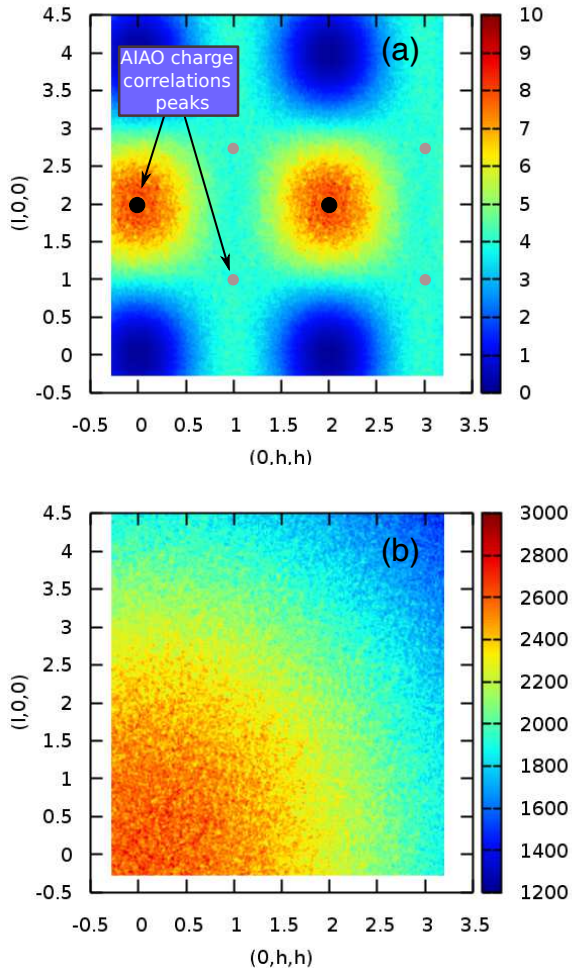}  
    \caption{(a) Cut in the $[lhh]$ plane of the charge-charge correlation function for the monopole liquid, calculated at $T = 0$ using the CMA (lattice size $L = 32$). Although, there are only non-interacting single monopoles present, the diffuse correlations around $[200]$ are a broad version of those for an AIAO arrangement (a crystal of double charges) which narrow peaks are indicated by spots with sizes growing with their intensities. (b) Simulated neutron structure factor for the same monopole liquid, in the same region of ${\bf k}$-space. It shows that ---like in a perfect paramagnet--- there are no spin correlations, something that is preserved at all temperatures for the BML. (To simplify the comparison with the spin ice Dy$_2$Ti$_2$O$_7$ we have used the magnetic form factor corresponding to Dy$^{+3}$)} 
    \label{corrmon}
\end{figure}

In order to check these results and to give support to the discussion on the next section on the emergence of spin short-range-order by disorder, we calculated the Fourier transform of the charge-charge correlation function \footnote{The charge-charge correlation function was calculated as if it were a system of real charges, using the corresponding topological charge sitting at the center of a tetrahedron as a variable.} and the neutron scattering structure factor of the spin system (for details, see Appendix \ref{SD}) for the ML ground state. The Monte Carlo simulations were performed at essentially $T = 0$ with the Conserved Monopoles Algorithm (CMA, see Appendix \ref{SD} and Reference~\citenum{Borzi2013}), for a lattice size $L = 32$. A color plot of the charge-charge correlation function is shown in Fig.~\ref{corrmon}(a) for the $[lhh]$ reciprocal space plane. As a reference, we have marked the ${\bf k}$-space values where narrow Bragg peaks occur (the size of the symbols indicating larger intensity) for the correlation function on a perfectly ordered monopole crystal with the Zn-blende structure. The coincidence of the centers of both sets of peaks confirms that we are in the presence of a liquid-like phase of charges, where effective interactions are not strong enough to turn it into a crystal. The spread of the diffuse peaks ($\Delta k > 1 \times 2\pi/a$) indicates that the correlation length is smaller than the cell parameter ($a$) of the conventional cubic unit cell.
On the other hand, the neutron scattering structure factor\footnote{In order to facilitate its comparison with experiments (see next section) we have used for the magnetic moments the form factor corresponding to  Dy$^{+3}$.} (i.e., the fourier transform of the spin-spin correlation function) shown in Fig.~\ref{corrmon}(b) is that of a paramagnet \cite{bramwell2011lacroix}. We believe this result is non-trivial: the geometry of the lattice and internal structure of the single monopoles are such that whatever energetics imposing a ground state of equiprobable 3in-1out or 1in-3out configurations will end in a perfect paramagnet at $T = 0$. So as to stress more this fact, note that the structure factor we obtained for the ML at $T=0$ K (Fig.~\ref{corrmon}(b)) should coincide with that calculated for Dy$_2$Ti$_2$O$_7$ for $T \rightarrow \infty$ .(We will soon discuss the temperature scale for this to happen).

\subsection{The \emph{Balanced} Monopole Liquid model (BML): simulations and exact solution at finite $T$}

To investigate the thermal properties of spin and charge correlations in the ML and its thermodynamics, we need to introduce a model. It is clear from Subsection~\ref{NNmodel} that it is not possible to obtain a liquid ---or even a crystal--- of single monopoles out of Eq.~\ref{effnn}, something that cannot be changed by the addition of dipolar interactions \cite{Melko2004,Guruciaga2014}. 
We generalize further this fact in Appendix A, showing that a ML cannot be stabilized in the pyrochlore lattice by the inclusion of any combination of translationally invariant \emph{pairwise} interactions. 
Previous works have resorted to magnetic fields \cite{Castelnovo2008,Guruciaga2016} ---cancelling the residual entropy of the system--- or to artificial constraints \cite{Borzi2013, BBartlett2014} in order to have fluids or crystals of single monopoles. We have found that a simple model Hamiltonian leading to a ML ground state can be written including terms with the product of the four pseudospins on the vertices of each tetrahedron $k$ of the pyrochlore lattice:
 \begin{equation}\label{eq1}
     {\cal H}_{BML}/k_B= \sum_k J_{\square}  \prod_i^4 \sigma^{k}_i,
 \end{equation}
 with $J_{\square}>0$, the strength of what we call the  balanced ML model. This Hamiltonian gaps the monopole configurations from those containing neutral tetrahedra or double monopoles (on equal footing), adding $2J_{\square}$ per excitation. Equation \ref{eq1} can be thought of as a variation of that proposed by Jaubert \cite{jaubert2015s} to study a crystal of single monopoles.  It is useful to think the BML as an effective spin model, the result of having integrated out the crystal lattice or the dipolar electric degrees of freedom \cite{jaubert2015s,jaubert2015p,Khomskii2012}. However, aside from the considerations in Appendix \ref{GB}, we make no attempt here to justify the precise form of our Hamiltonian, in favor of a discussion of its consequences and possible association with other systems and materials.

 Quite remarkably, the Hamiltonian in Eq. \ref{eq1} was %(apart from a trivial change in sign) 
 solved exactly for the pyrochlore lattice in the late eighties by Barry and Wu (see Ref. \citenum{barry1989}), in a context completely different from that of geometrical frustration and magnetic charges. The authors calculate the partition function of the system which in the thermodynamic limit turns out to be that of a system consisting of $N/2$ free dichotomic variables:
 \begin{equation}
 {\cal Z} = 2^{\frac{N}{2}} \left(2\cosh({\beta J_{\square}})\right)^{\frac{N}{2}}
 \end{equation}
 from which we can recover the residual entropy of the ML calculated above. 
 This results and the pair spin correlations calculated by Barry and Wu show that the BML extends our results for the ground state to all temperatures: the spin system described by the BML consists (at zero magnetic field) of a single paramagnetic phase with no spin correlations. Notably, the random thermal excitation of neutral and double charges in a ratio $6/2$ clearly dilutes the single monopoles, but do not affect the degree of spin correlation (and it is in this sense that we call this system a \emph{balanced} ML\footnote{Inverting the sign of $J_\square$ leads to a ground state populated only by neutral sites and double monopoles in proportion $6:2$; this ratio does not change when increasing temperature. We have checked that this is, as expected, \emph{another} perfect paramagnet at all temperatures.}).

We have studied the thermodynamics of the system by means of standard single spin flip Monte Carlo simulations (see Appendix \ref{SD}).  Fig.~\ref{fig1}(b) shows in the right axis that for $T/J_{\square} \lessapprox 1$ the \emph{single} monopole density ($\rho$) approaches 1 monopole per tetrahedron. As shown in the same figure, this density can be obtained from the exact solution for the energy~\cite{barry1989} and calculated as $\rho=1/2 (1 + \tanh(J/T))$. At the same temperatures, the specific heat is seen to peak ($C_V$, both simulated and calculated~\cite{barry1989}), marking the extinction of double monopoles and neutral excitations. We have obtained the entropy of the system (Fig.~\ref{fig1}(b), left axis) by integration of $C_V/T$, and imposing Pauling's estimate as the integration constant.

We checked that the monopole liquid gives the same isotropic pattern as in Fig.~\ref{corrmon}(b) when the temperature increases, reflecting that there are no spin-spin correlations \emph{at any temperature}.  Although it agrees with the exact results obtained in Ref.~\citenum{barry1989}, this seems to be hardly compatible with the measured increase in entropy (Fig.~\ref{fig1}(b)), but for the fact that excitations (in the form of double monopoles and 2in-2out tetrahedra) are now diluting the single monopoles in random positions. This idea can be quantitatively and beautifully tested by calculating the entropy of mixing that results from this monopole dilution. The curve shown in Fig.\ref{fig1} with blue symbols is the entropy increase because of mixing ($\Delta S_{mix}(T)/Nk_B = -1/2 \times [\rho(T) \ln(\rho(T)) + (1-\rho(T)) \ln(1-\rho(T))]$) to which we have added $S_{res}$. The excellent coincidence with the curve obtained from the specific heat is a nice way to emphasize that for the balanced monopole liquid model of Eq.~\ref{eq1} only the charge degrees of freedom are relevant.% \footnote{For spin ice $\Delta S_{mix}$ is a good approximation \emph{only} for $T < J_{nn}$ (i.e., for small monopoles concentration). As expected, the entropy of mixing cannot account for the entropy increase when the dipolar spin correlations start to weaken.}.
    
    %While this mixing of charges \footnote{Remarkably, $S_{mix}$ implies that only matters if a site is occupied by a single monopole or not --without the need to specify if the diluted site is neutral or with a double charge.} within the sites of the diamond lattice undoubtedly increases the global entropy of the spin system, it is a mystery to be explained later why the simultaneous inclusion of the two type of excitations in a ratio $6:2$ does not introduce (on average) any pair spin correlations.

%%%%%%%%%%%%%%%%%%%%%%%%%%%%%%%%%%%%%%%%%%%%%%%%%%%%%%%%%%%%%%%%%%%%%
%%%%%%%%%%%%%%%%%%%%% RESULTS & DISCUSSION %%%%%%%%%%%%%%%%%%%%%%%%%%
%%%%%%%%%%%%%%%%%%%%%%%%%%%%%%%%%%%%%%%%%%%%%%%%%%%%%%%%%%%%%%%%%%%%%
%%%%%%%%%%%%%%%%%%%%%%%%%%%%%%%%%%%%%%%%%%%%%%%%%%%%%%%%%%%%%%%%%%%%%

%\section{\texorpdfstring{$J_{\square}>0$}{TEXT}: the monopole liquid}

%%%%%%%%%%%%%%%%%%%  Revisar luego si hace falta lo de abajo %%%%%%%%%%%%%%%%%%%%%%%%%%%%%%%%%%%%%%%%%%%%%%%%%%%%
% \textcolo{blue}{We analyze now} the four spin monopole liquid model in terms of charge degrees of freedom. 

 %{\color{red} *** ojo que esto esta dicho de otro modo en la seccion anterior: NN charge correlations can be calculated by assuming a given charge is in a tetrahedron and counting all possible spin configurations for the four nearest neighbor tetrahedra. It is easy to check that each positive single monopole (topological charge $\sum_i^4 \sigma^{k}_i=+2$) is on average  surrounded by a total charge of $-2$. }

After having studied the BML given by Eq.~\ref{eq1}, the obvious question to ask is how different perturbations will break the delicate balance that makes the spins in this system uncorrelated at all temperatures. The important case of a magnetic field will be studied elsewhere \footnote{D. Slobinsky and R. A. Borzi, in preparation}; in the next section we consider other interactions, likely to be present in real pyrochlore materials.

%%%%%%%%%%%%%%%%%%%%%%%%%%%%%%%%%%%%%%%%%%%%%%%%%%%%%%%%%%%%%%%%%%%%%
%%%%%%%%%%%%%%%%%%%%%%     Interactions     %%%%%%%%%%%%%%%%%%%%%%%%%
%%%%%%%%%%%%%%%%%%%%%%%%%%%%%%%%%%%%%%%%%%%%%%%%%%%%%%%%%%%%%%%%%%%%%
%%%%%%%%%%%%%%%%%%%%%%%%%%%%%%%%%%%%%%%%%%%%%%%%%%%%%%%%%%%%%%%%%%%%%
\section{Effect of perturbations on the Balanced Monopole Hamiltonian \texorpdfstring{${\cal H}_{BML}$}{TEXT}}

The BML is only one particular model (with the peculiarities that it is exactly balanced to be a perfect paramagnet at all temperatures, and that it has an analytic solution) leading to a ML ground state. 
We will subject the BML to perturbations, and study their effect in two different aspects: $i-$ how, while preserving the ML ground state, new terms can imbalance the BML and thus affect correlations at finite temperature; $ii-$ how different perturbations could destabilize the ML. Regarding the first point, we will consider in the next two subsections the usual nearest neighbor spin interactions (Eq. \ref{effnn}) as a small perturbation to Eq. \ref{eq1}. For $J_{nn}$ small enough, the effective ferro or antiferromagnetic character of nearest neighbor interactions will determine the preeminence of one or other type of low energy excitations (neutral 2in-2out tetrahedra, or double monopoles, respectively) to the ML. This, in turn, will affect in very different ways the system on increasing temperature. In the third subsection we explore the possibility of changing the ML ground state by adding interactions.

%%%%%%%%%%%%%%%%%%%%%%%%%%%%%%%%%%%%%%%%%%%%%%%%%%%%%%%%%%%%%%%%%%%%%%%%%%%%%%%%%%%%%%%%%%%%%%%%%%%%%%%%%%%%%%%%%%%%%%%%%%%
\subsection{ML ground state with excess 2in-2out excitations: \texorpdfstring{$J_{nn} > 0$}{TEXT}}

\begin{figure}
     \includegraphics[width=0.45\textwidth]{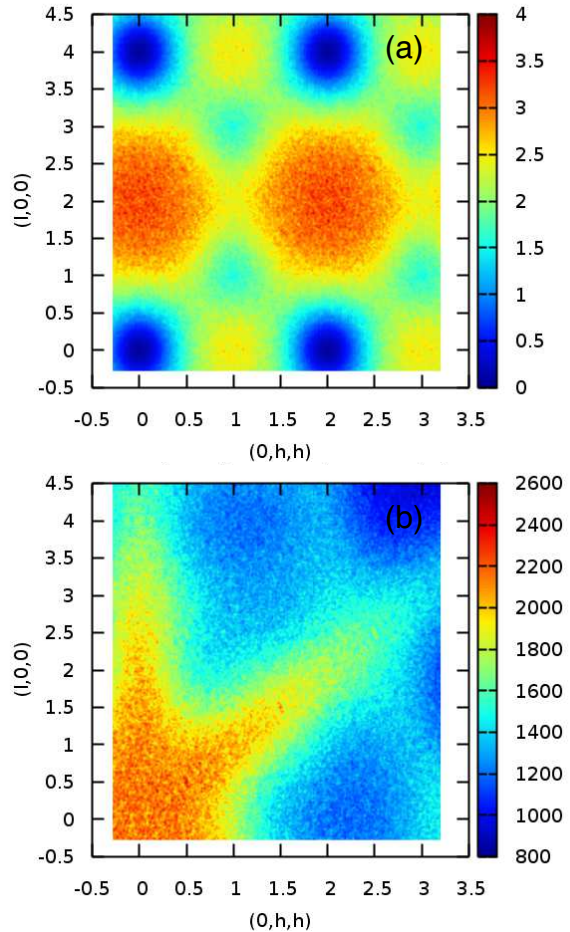}  
    \caption{(a) Charge-charge correlation function for a monopole liquid perturbed by a nearest neighbors interaction term favoring neutral sites ($L = 32$). The perturbation parameter $J_{nn}$ and temperature are such that any neutral site or single monopole can occur with equal probability, with a negligible concentration of double charges. (b) Simulated neutron diffuse scattering intensities for the same configurations. The charge correlations in panel (a) look broader than those for the ML (Fig.~ \ref{corrmon}(a)), with a magnitude reduced approx. by half. On the other hand, anisotropic spin correlations have now developed in panel (b), and present a structure comparable to that of a spin ice material at high temperature. Remarkably, dipolar-like spin correlations (absent at $T=0$) develop in our system with \emph{increasing} temperature.} 
    \label{jge0}
\end{figure}

We consider now adding to Eq. \ref{eq1} a nearest neighbors interaction term favoring the ice rule (Eq. \ref{effnn}). For moderate values of $J_{nn}$, its effect is that of raising the probability of finding a neutral tetrahedron (2in-2out) with increasing temperature. %\footnote{As discussed in Section~\ref{NNmodel}, Eq. \ref{effnn} is only sensitive to the type of charge in a given tetrahedron, effectively assigning energies  $\epsilon_n < \epsilon_s < \epsilon_d$ for neutral, single or double charges for positive $J_{nn}$. Following their proportionality with $J_{nn}$, the inverse energy order would be attained for the antiferromagnetic case.
In order to show how this perturbation affects the system we consider an extreme case\footnote{A less extreme case, which may include a lower density of neutral sites or the combination of neutral and double monopoles, leads to a much smaller effect in the same direction.}: $T$ and $J_{nn}$ are such that all single monopoles and 2in-2out configurations have the same probability to manifest ($1/(8+6)$), with a negligible density of double monopoles. Fig.~\ref{jge0}(a) shows the charge-charge correlation function simulated in these conditions. The widening and shortening of the peaks indicate that charge correlations decrease as single monopoles are diluted by neutral sites, pushing it towards a gas phase. Just the opposite occurs on the spin channel (Fig.~\ref{jge0}(b)) where the addition of 2in-2out sites has transformed the spherically symmetric pattern into an anisotropic one. Indeed, the pattern resembles that measured \cite{Fennell2009,Chang2010} or simulated (see Fig.~\ref{appSI} in Appendix \ref{SD}) for spin ice at high temperature, with diffuse scattering more noticeable for ${\bf k}$ along $[l00]$ and $[lll]$. We are then faced with a quite peculiar situation. Starting from a perfect spin paramagnet with charge correlations at $T = 0$, we found that the addition of neutral sites diminishes these correlations, while on the other hand it enhances \emph{spin} correlations, moving the system towards a spin liquid. Given that the spin system does not reach long range order (or, more appropriate, the power law correlations typical of the emergent gauge field in spin ice \cite{Henley2010,Castelnovo2011}) we can term this `short range ({\it topological}) order by disorder' \footnote{We will see later that this could not have happened starting from the antiferromagnetic AIAO ground state, since double monopoles do not allow for the propagation of dipolar correlations.}.

As discussed in the introduction, it was found in previous experimental \cite{Chang2010} and numerical \cite{Isakov2005} studies that spin ice correlations can be seen up to temperatures much higher than the typical energy scale ($J_{nn}$), and even in the presence of strong dilution \cite{Chang2010}.
This important issue was addressed before by Sen {\it et al.}~\cite{Sen2013}, and is also implicit on Ref.~\citenum{bramwell2017harmonic}. Both approaches are based on a charge approximation in which double monopoles were neglected. We will argue below that the absence of double monopoles is indeed crucial to obtain dipolar correlations at high temperatures. A first step in this direction was taken by Brooks-Bartlett {\it et al.} \cite{BBartlett2014} who noted that single monopoles of fixed charge can have associated a fluctuating moment (a prerequisite to have a Coulomb phase~\cite{Henley2010}), something that is not possible for double charges. However, as made evident by the balanced ML case ---where we observed a total absence of correlations in a highly fluctuating scenario (Fig. \ref{corrmon}(b))--- the existence of fluctuating magnetic moments is not a sufficient condition.

Complementing these previous studies we can understand now the persistence of correlations at high temperature in a simple manner (without taking into account dipolar interactions) by analyzing spin correlations for the present case  in which all neutral and single monopole configurations are equally probable. 
We proceed as before, by checking the bias imposed by a given spin on any other in the same tetrahedra. We should now count the six possible 2in-2out configurations in addition to the eight monopole ones. If the absence of double monopoles is guaranteed, the fixing of a spin direction turns now the probability of having a pseudospin in the same tetrahedron pointing \emph{in the opposite direction} ({\it i.e.}, having the opposite sign) $1/7$ more likely than otherwise. This is of course much smaller than the probability of $1/3$ that we got for spin ice, but it is not negligible, and is responsible for dipolar-like correlations in the former perfect paramagnet. From the point of view of spin ice we can say that, basically, monopoles dilute the dipolar correlations propagated by neutral sites, without erasing them.

We will now point to the evident fact that raising the temperature to infinity should restore the isotropic correlations of a perfect paramagnet, a phase which (remarkably) is now at \emph{both ends of the temperature axis}. The expected destruction of spin correlations at high temperatures can be made less trivial by inferring that it is then the presence of the double monopoles what makes all the difference. We can now add to this that ---even in the presence of strong local moment fluctuations--- only when the ratio of neutral tetrahedra and double monopoles is near $6:2$ the dipolar-like spin correlations will be erased.  A good example of this is the BML at all temperatures, but also the ML (where there are no double monopoles or neutral cites) which can be thought as the BML in the limit $T \rightarrow 0$. An important instance with an experimental counterpart is that of spin ices for $T \gg 8J_{nn}$ (with $J_{nn} \approx 1\ K$).
At $T = 10 K$ ($T \approx 8J_{nn}$), where experiments still show a spin ice-like neutron pattern in Ho-titanate samples \cite{Chang2010}, our simulations using Eq.~\ref{effnn} show that the fraction of single monopoles per tetrahedron is near $1/2$, but the ratio of neutral to double monopole sites is bigger than $7$ (see Appendix \ref{SD} for the magnetic structure factor we obtained). As shown in Ref.~\citenum{Sen2013}, the inclusion of dipolar interactions (present in real materials, but not considered here) should enhance even more these correlations, and contribute to define better the pinch points in Fig.~\ref{jge0}(b) and Fig.~\ref{appSI}. 

%This is the reason why dipolar-like correlations in spin ice systems survives to very high temperatures: the energetically cheaper single monopoles preserve in some degree these correlations.} As long as double monopoles are not present in significant proportions (their high temperature limit is $2/16$) spin ice correlations would persist. At $T = 10 K$, where experiments still show a spin ice-like neutron pattern even in the presence of disorder \cite{Chang2010}, our numerical simulations on spin ice show that the fraction of single monopoles is very near $1/2$, but that of doubles is less than $1/16$ (see Appendix \ref{SD}).

%\subsection{Correlations due to double monopole excitations: \texorpdfstring{$J_{nn} < 0$}{TEXT}}
\subsection{ML ground state with excess double monopole excitations: \texorpdfstring{$J_{nn} < 0$}{TEXT}}

\begin{figure}
    \includegraphics[width=0.45\textwidth]{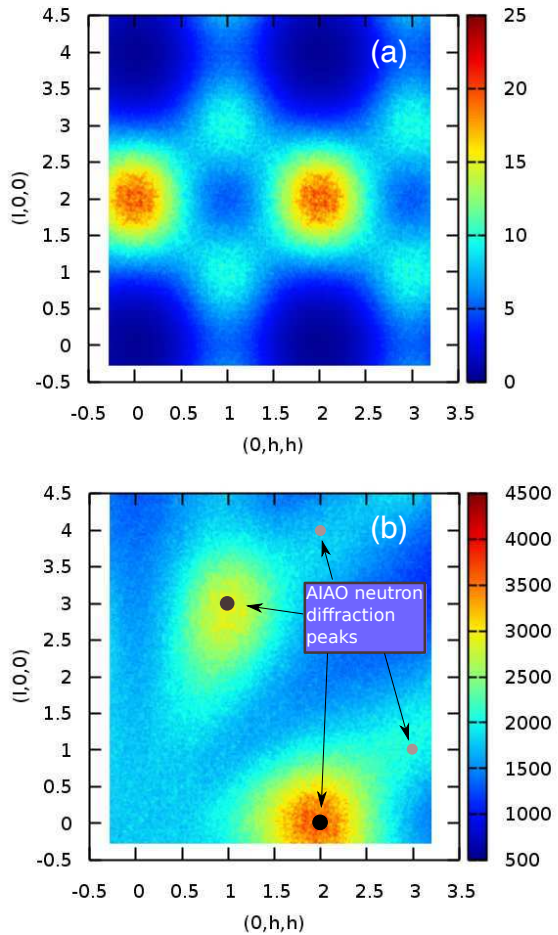}  
    \caption{(a) Charge-charge correlation function for a monopole liquid perturbed by a nearest neighbors interaction term favoring `all-in' or `all-out' configurations ($L=32$). The parameters and temperature were chosen such that any monopole configuration among the ten possible ones (double or simple) can occur with equal probability, with a negligible concentration of neutral sites.  Now the charge correlations look much sharper than those for the ML (Fig.~\ref{corrmon}), and with extra intensity. (b) Simulated neutron diffuse scattering intensities for the same configurations. We observe again the development of anisotropic spin correlations. Now, a relatively small concentration of thermally generated double monopoles favors the appearance of broad peaks in positions compatible with the crystal of monopoles, the all-in/all-out antiferromagnet (marked with circles, their sizes indicating their intensities), in a case of classical `\emph{short-range} order by disorder'.}
    \label{jlt0} 
\end{figure}

When the added perturbation term (Eq.~\ref{eq1}) favors AIAO order, its effect at finite temperature is to increase the density of double monopoles. It has been previously shown that introducing this type of excitations in a disordered monopole ground state can lead to a crystal of monopoles \cite{Guruciaga2016}, in an example of classical order by disorder \cite{villain1980,Lacroix2011}. One important difference between this case and the ML we are now studying is the huge residual degeneracy present in our system. In order to test the occurrence or not of this phenomenon we study again the extreme case, with a value for $T$ and the perturbation such that any single or double monopole configuration has the same chance to manifest ($1/10$), with negligible proportion of neutral tetrahedra. Fig.~\ref{jlt0}(a) shows that the charge-charge correlation function has higher and narrower peaks than the BML ---showing that monopole correlations are stronger, with a correlation length extending approximately over a unit cell--- but that crystallization has not taken place. 

Regarding the spin channel, we can see in Fig.~\ref{jlt0}(b) that the perfect paramagnet has developed correlations in the form of diffuse scattering around $[311]$ and $[022]$, the same positions of the perfectly ordered AIAO state. In this case then, both the monopole and spin channels are examples of short range order by disorder.  Proceeding as we did before, we can evaluate how the double monopoles occurring with probability $2/10$ unbalance the correlation of a nearest neighbor spin within a single tetrahedron. We find that once we fix the direction of a pseudospin, the probability that a second pseudospin in the tetrahedron points \emph{in the same direction} is $1/5$ larger than otherwise. The change in sign in the average pseudospin value explains the qualitative differences observed in the correlation functions (Fig.~\ref{jge0}(b) and Fig.~\ref{jlt0}(b)).

\subsection{Effect of other interactions on the ML ground state.}
Eqs.~\ref{effnn} and~\ref{eq1} have the particularity of operating effectively at a charge level, where $J_{nn}$ and $J_\square$ function like chemical potentials for the creation of monopoles or neutral excitations. In general, spin interactions cannot be interpreted fully at this level, and they will tend to break the massive accidental degeneracy of the ML in favor of certain spin configurations (with their associated charge arrangement) at low temperatures. Long range dipolar interactions are a good example of these, since they can be thought as having a dominant effective charge-charge Coulomb component, together with a spin-spin correction~\cite{Castelnovo2008} that decreases with monopole distance $r$ like $1/r^5$. The addition of dipolar interactions to Eq.~\ref{eq1} will then make the ML crystallize into a zincblende crystal of monopoles on decreasing temperature, in a single or a series of steps (depending on the relative strength of ${\cal H}_{BML}$). These may involve first a solid of single monopoles ---a Coulomb spin-liquid studied in Refs.~\citenum{BBartlett2014} and~\citenum{jaubert2015s}--- and then either freeze the fragmented spin liquid~\cite{BBartlett2014} into one of the ground states previously determined \cite{Borzi2013}, or transform it into a crystal of double charges.

While charge-like interactions can generally be more intuitively handled, it would be impossible to list the effect of different spin-spin interactions (with varying range and strength) on the ML. It is then a pleasant surprise that second and third nearest neighbor exchange interactions could in principle be tuned to preserve the charge-charge character of Eq.~\ref{eq1}. Their inclusion is not a mere detail as would be naively guessed. If attraction between like monopoles is triggered \cite{Udagawa2016,Rau2016} it can open the door to very sophisticated charge disordered phases different from the ML \cite{Rau2016} ---for example based on \emph{jelly fish} structures. Aside from thermodynamics, these structures may have a strong influence on the dynamics of the system, something that has been studied in relation with the spontaneous Hall effect observed in Pr$_2$Ir$_2$O$_7$ \cite{Udagawa2016,machida2007unconventional}. Extra interactions, a magnetic field, quantum effects, and the interplay with other degrees of freedom, could lead to more sophisticated single monopole crystals, as the double monopole layer seen in Tb$_2$Ti$_2$O$_7$ \cite{Sazonov2012,BBartlett2014,jaubert2015p}. 
Indeed, the physics found in the titanate of Tb \cite{Fennell2014,Sazonov2012,BBartlett2014}, based on a very low $J_{nn}$ and a strong coupling between magnetic and elastic or dipolar electric degrees of freedom \cite{jaubert2015s,jaubert2015p,Khomskii2012}, bears some qualitative resemblance with the physics behind the ML, suggesting a promising route for its realization.

\section{Conclusions}

One of the contributions of this paper has been the description of a new state of monopole matter. We studied a monopole liquid in the pyrochlore lattice, a charge-disordered phase with high density of single monopoles, in the absence of an applied magnetic field. We found the residual entropy per spin of this phase to be $\ln(2)/2$ (amounting to one free Ising variable per tetrahedron) a result which is exact in the thermodynamic limit.

Although we named this phase after its magnetic charge properties, the monopole liquid turned out to be an ambivalent phase. While at $T = 0$ it is a liquid from the point of view of monopoles, from that of spins (as it would be seen in diffuse neutron scattering measurements) it is a `spin gas' ({\it i.e.}, a perfect paramagnet) free of spin-spin correlations.

After proving that a ground state of single monopoles cannot be stabilized by translationally invariant pairwise spin interactions of any range, we proposed a four spin Hamiltonian, and we use it to study the monopole liquid thermodynamics and to measure charge as well as spin-spin space correlations. We name this model the \emph{balanced} monopole liquid (BML): the thermal creation of neutral and double monopole excitations in proportion $6:2$ does not perturb the spin correlations, making the BML a perfect paramagnet at any given temperature. Indeed, thermal disorder is directly related with the entropy of mixing that results from the dilution of single monopoles with neutral and double monopoles. 

The balance characterizing the complete lack of spin-spin correlations in the BML model is affected by small perturbations, and we learned new physics from their inclusion. The main consequence of adding moderate nearest neighbors ferro or antiferromagnetic interactions is changing the proportion of excitations of each type. As in the phenomenon of classical order by disorder, thermal excitations tend to imbalance the magnetically disordered BML towards two different forms of order.

The case in which the perturbation is ferromagnetic promotes neutral excitations; by diluting the single monopoles their effect is to reduce their charge correlations. On the other hand, they create spin correlations resembling those measured in spin ice materials and systems. For this one system, the short range order created by thermal disorder resembles that related with an emergent gauge field. This phenomenon lead us by another route~\cite{BBartlett2014,Sen2013} to show how the presence of strong correlations in spin ice materials at temperatures much higher than $J_{nn}$ can be understood by the relative low concentration of double charges. The absence of spin correlations in the strongly fluctuating environment of the monopole liquid (in the limit $T \rightarrow 0$), spin ices for $T \gg 8J_{nn}$, and the balanced monopole liquid (at all temperatures) can be rationalized as cases in which the fraction of neutral sites to double monopoles occurs in a ratio approaching $6:2$.
%This was further developed by showing that the system which is the converse of the monopole liquid (with a ground state populated by neutral sites as well as double monopoles, with single monopoles as excitations) is also a perfect paramagnet.
The antiferromagnetic perturbation ---favoring a majority of double monopoles as excitations, which are now in a ratio of less than $6:2$ respect to neutral sites--- is the converse of the previous case. Charge-charge correlations increase in strength by the insertion of double monopoles, while spin correlations generate a diffuse version of the crystal of double charges (the all-in/all-out state) in a sort of short range order by disorder.

We think that, in the same way we have used this system to understand in a new light spin ices, the monopole liquid can be the starting point to rationalize the physics behind Tb$_2$Ti$_2$O$_7$, materials showing magnetic moment fragmentation, and ---more generally--- of \emph{monopole matter}\cite{Morris2009,Sazonov2012,Borzi2013,BBartlett2014,Udagawa2016,Petit2012,jaubert2015p,Petit2016,Lefrancois2017}.

%%%%%%%%%%%%%%%%%%%%%%%%%%%%%%%%%%%%%%%%%%%%%%%%%%%%%%%%%%%%%%%%%%%%%
%%%%%%%%%%%%%%%%%%%%%%% Acknowledgments %%%%%%%%%%%%%%%%%%%%%%%%%%%%%
%%%%%%%%%%%%%%%%%%%%%%%%%%%%%%%%%%%%%%%%%%%%%%%%%%%%%%%%%%%%%%%%%%%%%
%%%%%%%%%%%%%%%%%%%%%%%%%%%%%%%%%%%%%%%%%%%%%%%%%%%%%%%%%%%%%%%%%%%%%
\begin{acknowledgments}
The authors wish to thank  R. Moessner, B. Schmidt, D. Rosales, A. P. Mackenzie, S. A. Grigera, C. Hooley, and P. Holdsworth for useful discussions, and acknowledge financial support from ANPCyT (PICT 2013 N$^{\circ}$2004, PICT 2014 N$^{\circ}$2618), and CONICET (PIP 0446). They also want to thank the peer reviewers for their useful comments and drawing their attention to Ref. \citenum{barry1989}.
%We thank one of the referees for noticing that Eq.~\ref{Eq1} has been exactly solved in ref. xx.????????????
\end{acknowledgments}

%%%%%%%%%%%%%%%%%%%%%%%%%%%%%%%%%%%%%%%%%%%%%%%%%%%%%%%%%%%%%%%%%%%%%
%%%%%%%%%%%%%%%%%%%%%%%%%% Appendices %%%%%%%%%%%%%%%%%%%%%%%%%%%%%%%
%%%%%%%%%%%%%%%%%%%%%%%%%%%%%%%%%%%%%%%%%%%%%%%%%%%%%%%%%%%%%%%%%%%%%
%%%%%%%%%%%%%%%%%%%%%%%%%%%%%%%%%%%%%%%%%%%%%%%%%%%%%%%%%%%%%%%%%%%%%
\appendix

%%%%%%%%%%%%%%%%%%%%%%%%%%%%%%%%%%%%%%%%%%%%%%%%%%%%%%%%%%%%%%%%%%%%%
%%%%%%%%%% No monopoles from nn pairwise interaction %%%%%%%%%%%%%%%%
%%%%%%%%%%%%%%%%%%%%%%%%%%%%%%%%%%%%%%%%%%%%%%%%%%%%%%%%%%%%%%%%%%%%%
%%%%%%%%%%%%%%%%%%%%%%%%%%%%%%%%%%%%%%%%%%%%%%%%%%%%%%%%%%%%%%%%%%%%%
\section{Impossibility to generate a monopole liquid from translationally invariant static pairwise interactions}\label{GB}

Some of the results shown along this paper have been derived from a four spin Hamiltonian (Eq.~\ref{eq1}). As a way to justify the need for such a Hamiltonian, we prove in this Appendix that a ML ground state can not be stabilized by translationally invariant \emph{pairwise} spin interactions, irrespectively of their range.

We will prove it by contradiction, assuming first that the ML ground state can be obtained using a set of pairwise interaction constants $J_{ij}$ depending only on the relative position of pseudospins $\sigma_i$ and $\sigma_j$ on the pyrochlore lattice sites $i$ and $j$. $J_{ij}$ should be such that all spin configurations with one single monopole per tetrahedron have the same energy $E_0$, with $E_0$ smaller than the energy of any other configuration. It will probe useful to write the energy $E$ of a given configuration as the sum of two terms:
\begin{equation} 
E= \sum_{k,l,a,b} J_{kl}^{ab} \sigma_k^a \sigma_l^b = D + T
\end{equation}
in which now $k$ and $l$ indicate the up tetrahedra to which the pseudospins belong, $a$ and $b$ correspond to one of the four spin types in a given tetrahedron. Implicitly, self-energy terms in which $a=b$ and $k=l$ simultaneously, have been removed from this and all following summations. $D \equiv \sum_{k,l, a = b} J_{kl}^{aa} \sigma_k^a \sigma_l^a$ and $T \equiv \sum_{k,l, a\neq b} J_{kl}^{ab} \sigma_k^a \sigma_l^b $ are the diagonal and off-diagonal terms on the type of spin.

The \textit{reductio ad absurdum} is reached by computing the energy $E$ for three different spin configurations which we chose to be perfectly ordered: $i-$ a crystal of double monopoles (which we label AIAO); $ii-$ a crystal of single monopoles, which belongs to the monopole liquid phase and has minimum energy $E_0$ (ML); $iii-$ a spin ice configuration with maximum magnetization along the $[100]$ direction (SI). In all three cases, the whole spin configuration can be assembled by repeating in every up tetrahedron of the pyrochlore lattice a fixed spin arrangement (see  Table~\ref{apptab1}).

\begin{table}
    \centering
    \begin{tabular}{c|cccc}
            & 1 & 2 & 3 & 4 \\
    \hline
    AIAO    & + & + & + & + \\
    ML      & + & + & + & - \\
    SI      & + & + & - & - \\
    \hline
    \end{tabular}
    \caption{Basis for the configurations of double (AIAO) and single monopole crystal (ML), and spin ice (SI) configurations. The spin configurations are obtained by repeating these arrangements over all up tetrahedra of the pyrochlore lattice.}
    \label{apptab1}
\end{table}

Given their regularity, the diagonal term $D$ is the same in all three configurations, leaving only the off-diagonal term $T$ to be computed. We can now separate the off-diagonal term of the AIAO configuration into six contributions:
\begin{equation}\label{appTaiao}
    T_{AIAO} = \alpha^{12} + \alpha^{13} + \alpha^{14} + \alpha^{23} + \alpha^{24} + \alpha^{34}   
\end{equation}
where $\alpha^{ab} =\sum_{k,l,a\neq b} J_{kl}^{ab}$ (the signs in Eq. \ref{appTaiao} come from the fact that $\sigma_i^a \sigma_j^b=1$ for all $a\neq b$ for this configuration). We can obtain similar equations for the \textit{ML} and \textit{SI} configurations using Table~\ref{apptab1}:
\begin{eqnarray}
     T_{ML} = \alpha^{12} + \alpha^{13} - \alpha^{14} + \alpha^{23} - \alpha^{24} - \alpha^{34}   \\
     T_{SI} = \alpha^{12} - \alpha^{13} - \alpha^{14} - \alpha^{23} - \alpha^{24} + \alpha^{34}   
\end{eqnarray}

Symmetry considerations guarantee that all $\alpha^{ab}$ are equal ($\alpha^{ab}=\alpha$) leading to $T_{AIAO}=6\alpha$, $T_{ML}=0$ and $T_{SI}=-2\alpha$. 
The absurd is reached by noting that $E_0 = D < D+6\alpha$ and $E_0 = D < D-2\alpha$.

This contradiction demonstrates that it is impossible to obtain a monopole liquid from pairwise translationally invariant interactions, paving the way for the search of the ML within many body interactions (as we did in the text) or by dynamic models including further degrees of freedom \cite{Slobinskyinprep,jaubert2015s,jaubert2015p}.

%%%%%%%%%%%%%%%%%%%%%%%%%%%%%%%%%%%%%%%%%%%%%%%%%%%%%%%%%%%%%%%%%%%%%
%%%%%%%%%%%%%%%%%%%%%%%%%%%% Methods %%%%%%%%%%%%%%%%%%%%%%%%%%%%%%%%
%%%%%%%%%%%%%%%%%%%%%%%%%%%%%%%%%%%%%%%%%%%%%%%%%%%%%%%%%%%%%%%%%%%%%
%%%%%%%%%%%%%%%%%%%%%%%%%%%%%%%%%%%%%%%%%%%%%%%%%%%%%%%%%%%%%%%%%%%%%
\section{Simulation details}\label{SD}
We provide here some details on the simulations used in the main text to study the equilibrium properties of the Ising pyrochlores. We simulated $L\times L \times L$ conventional cubic cells of the pyrochlore lattice ($16\times L\times L \times L$ spins) with Metropolis and Conserved Monopoles \cite{Borzi2013} algorithms. In all cases the boundary conditions were set to be periodic along the cubic primitive vectors.

\subsection{Metropolis algorithm}
In order to obtain Fig.~\ref{fig1} in the main text we used the Metropolis algorithm with the Hamiltonian in Eq. \ref{eq1} (main text) with the usual spin-flip dynamics and $L=8$. After reaching equilibrium, we averaged the data over $10-150$ independent runs and $5 \times 10^4 - 1.5 \times 10^6$ time-steps depending on the temperature.
    
\subsection{Conserved Monopoles Algorithm (CMA)}
We have used the conserved monopoles algorithm \cite{Borzi2013} to speed up simulations without leaving the monopole liquid phase. The situation is quite parallel to its use in spin ice within the two-in---two-out manifold \cite{Borzi2013,Xie2015,Baez16}. The usual Metropolis dynamics is modified to control the temperature and the density of magnetic charge independently. We start the simulation from a spin configuration in which all tetrahedra are occupied by single monopoles excluding two, neutral ones. From this initial state, the only spin flips allowed are those that leave the number of monopoles unchanged, producing then an effective movement of the neutral sites. The inclusion of these neutral tetrahedra has no measurable effect for large system sizes.

\begin{figure}
    \includegraphics[width=0.45\textwidth]{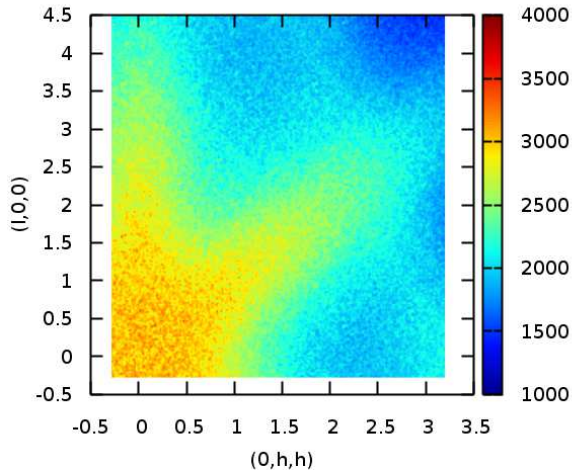}  
    \caption{Simulated spin structure factor for the nearest neighbors spin ice model of Eq.~\ref{effnn} ($L = 32$, $J_{nn} = 1.1$ K) at $T = 10$ K. The anisotropy observed on the pattern (characteristic of the Coulomb phase) can be seen at temperatures ten times  higher than  $J_{nn}$. At these temperature, the fraction of single monopoles per tetrahedron is near its high temperature limit of $1/2$, but the ratio of neutral to double  monopole  sites exceeds $7$, much bigger than the value of $3$ at which dipolar correlations disappear.} 
    \label{appSI}
\end{figure}

\subsection{Calculation of the charge correlation function}
The charge-charge correlation function has been calculated as:
\begin{equation}
    I(\boldsymbol{k})=\frac{2}{N}\sum_{ij}\langle Q_i Q_j\rangle\,e^{i\boldsymbol{k}\cdot \boldsymbol{r}_{ij}}
\end{equation}
%\textcolor{red}{Va o no va el 2/N?}

where $\langle ... \rangle$ indicates thermal average, $N/2$ is the number of tetrahedra, $Q_i$ represents the topological charge at position $\boldsymbol{r}_i$, and ${\boldsymbol{r}}_{ij}$ is the distance between monopoles. The sum runs over the charge positions on the vertices of the lattice `dual' to the pyrochlore, which is a diamond lattice.
All plots have been obtained in thermal averages over sets composed of $\approx 100$ configurations.

\subsection{Calculation of neutron diffraction structure factors}
The simulated neutron structure factors have been calculated following the expression:

\begin{equation}
    I(\boldsymbol{k})=\frac{[f(\left|\boldsymbol{k}\right|)]^2}{N}\sum_{ij}\langle\sigma_i\sigma_j\rangle\,\left(\boldsymbol{\mu}^{\perp}_i\cdot\boldsymbol{\mu}^{\perp}_j\right)\,e^{i\boldsymbol{k}\cdot \boldsymbol{r}_{ij}}
\end{equation}

\noindent where the sum now sweeps the pyrochlore lattices, $N$ is the number of sites, $\langle \sigma_i\,\sigma_j\rangle$ is the thermal average of the correlation between pseudospins at sites $i,j$; $\boldsymbol{\mu}^{\perp}_i$ is the component of the quantization direction of the spin $\boldsymbol{S_i} = \sigma_i \hat{\mu_i}$ at site $i$ perpendicular to the scattering wave vector $\boldsymbol{k}$: 

%%****************  REVISAR ESTO QUE CAMBIE SIMBOLOS ******************** Revisado 16/01/2018 - Demian
\begin{equation}
    \boldsymbol{\mu}^{\perp}_i=\hat{\mu}_i-\left(\hat{\mu}_i\cdot\frac{\boldsymbol{k}}{|\boldsymbol{k}|}\right)  \frac{\boldsymbol{k}}{|\boldsymbol{k}|}
\end{equation}

\noindent and $f(|\boldsymbol{k}|)$ is the magnetic form factor. We have chosen the magnetic factor of $Dy^{+3}$ as a benchmark to plot all structure factors. This factor is:
\begin{equation}
    f(|\boldsymbol{k}|)=c+\sum_{i=1}^4\,a_i\,\exp\left(-b_i\frac{\left|\boldsymbol{k}\right|^2}{16\pi^2}\right)
\end{equation}
\noindent with $a_i,b_i$ obtained from \footnote{\url{http://lamp.tu-graz.ac.at/~hadley/ss1/crystaldiffraction/atomicformfactors/formfactors.php}}. An example of the intensity as would be observed by neutron scattering could be seen in Fig. \ref{appSI}, in which we have plotted the structure factor for a nearest neighbors spin ice model (Eq.~\ref{effnn}, with $J_{nn}=1.1$ K) at $10$ K. We can see that even at such high temperatures there is still a marked anisotropy, with stronger scattering in the same directions of reciprocal space as has been measured at low temperatures.

%%%%%%%%%%%%%%%%%%%%%%%%%%%%%%%%%%%%%%%%%%%%%%%%%%%%%%%%%%%%%%%%%%%%%
%%%%%%%%%%%%%%%%%%%%%%%%%%%%%%%%%%%%%%%%%%%%%%%%%%%%%%%%%%%%%%%%%%%%%
%%%%%%%%%%%%%%%%%%%%%%%%%%%%%%%%%%%%%%%%%%%%%%%%%%%%%%%%%%%%%%%%%%%%%

\bibliography{references.bib}

%merlin.mbs apsrev4-1.bst 2010-07-25 4.21a (PWD, AO, DPC) hacked
%Control: key (0)
%Control: author (8) initials jnrlst
%Control: editor formatted (1) identically to author
%Control: production of article title (-1) disabled
%Control: page (0) single
%Control: year (1) truncated
%Control: production of eprint (0) enabled
\begin{thebibliography}{65}%
\makeatletter
\providecommand \@ifxundefined [1]{%
 \@ifx{#1\undefined}
}%
\providecommand \@ifnum [1]{%
 \ifnum #1\expandafter \@firstoftwo
 \else \expandafter \@secondoftwo
 \fi
}%
\providecommand \@ifx [1]{%
 \ifx #1\expandafter \@firstoftwo
 \else \expandafter \@secondoftwo
 \fi
}%
\providecommand \natexlab [1]{#1}%
\providecommand \enquote  [1]{``#1''}%
\providecommand \bibnamefont  [1]{#1}%
\providecommand \bibfnamefont [1]{#1}%
\providecommand \citenamefont [1]{#1}%
\providecommand \href@noop [0]{\@secondoftwo}%
\providecommand \href [0]{\begingroup \@sanitize@url \@href}%
\providecommand \@href[1]{\@@startlink{#1}\@@href}%
\providecommand \@@href[1]{\endgroup#1\@@endlink}%
\providecommand \@sanitize@url [0]{\catcode `\\12\catcode `\$12\catcode
  `\&12\catcode `\#12\catcode `\^12\catcode `\_12\catcode `\%12\relax}%
\providecommand \@@startlink[1]{}%
\providecommand \@@endlink[0]{}%
\providecommand \url  [0]{\begingroup\@sanitize@url \@url }%
\providecommand \@url [1]{\endgroup\@href {#1}{\urlprefix }}%
\providecommand \urlprefix  [0]{URL }%
\providecommand \Eprint [0]{\href }%
\providecommand \doibase [0]{http://dx.doi.org/}%
\providecommand \selectlanguage [0]{\@gobble}%
\providecommand \bibinfo  [0]{\@secondoftwo}%
\providecommand \bibfield  [0]{\@secondoftwo}%
\providecommand \translation [1]{[#1]}%
\providecommand \BibitemOpen [0]{}%
\providecommand \bibitemStop [0]{}%
\providecommand \bibitemNoStop [0]{.\EOS\space}%
\providecommand \EOS [0]{\spacefactor3000\relax}%
\providecommand \BibitemShut  [1]{\csname bibitem#1\endcsname}%
\let\auto@bib@innerbib\@empty
%</preamble>
\bibitem [{\citenamefont {Castelnovo}\ \emph {et~al.}(2008)\citenamefont
  {Castelnovo}, \citenamefont {Moessner},\ and\ \citenamefont
  {Sondhi}}]{Castelnovo2008}%
  \BibitemOpen
  \bibfield  {author} {\bibinfo {author} {\bibfnamefont {C.}~\bibnamefont
  {Castelnovo}}, \bibinfo {author} {\bibfnamefont {R.}~\bibnamefont
  {Moessner}}, \ and\ \bibinfo {author} {\bibfnamefont {S.~L.}\ \bibnamefont
  {Sondhi}},\ }\href {\doibase 10.1038/nature06433} {\bibfield  {journal}
  {\bibinfo  {journal} {Nature Physics}\ }\textbf {\bibinfo {volume} {451}},\
  \bibinfo {pages} {42} (\bibinfo {year} {2008})}\BibitemShut {NoStop}%
\bibitem [{\citenamefont {Ramirez}\ \emph {et~al.}(1999)\citenamefont
  {Ramirez}, \citenamefont {Hayashi}, \citenamefont {Cava}, \citenamefont
  {Siddharthan},\ and\ \citenamefont {Shastry}}]{Ramirez1999}%
  \BibitemOpen
  \bibfield  {author} {\bibinfo {author} {\bibfnamefont {A.~P.}\ \bibnamefont
  {Ramirez}}, \bibinfo {author} {\bibfnamefont {A.}~\bibnamefont {Hayashi}},
  \bibinfo {author} {\bibfnamefont {R.}~\bibnamefont {Cava}}, \bibinfo {author}
  {\bibfnamefont {R.}~\bibnamefont {Siddharthan}}, \ and\ \bibinfo {author}
  {\bibfnamefont {B.}~\bibnamefont {Shastry}},\ }\href@noop {} {\bibfield
  {journal} {\bibinfo  {journal} {Nature}\ }\textbf {\bibinfo {volume} {399}},\
  \bibinfo {pages} {333} (\bibinfo {year} {1999})}\BibitemShut {NoStop}%
\bibitem [{\citenamefont {Bramwell}\ and\ \citenamefont
  {Gingras}(2001)}]{Bramwell2001}%
  \BibitemOpen
  \bibfield  {author} {\bibinfo {author} {\bibfnamefont {S.~T.}\ \bibnamefont
  {Bramwell}}\ and\ \bibinfo {author} {\bibfnamefont {M.~J.}\ \bibnamefont
  {Gingras}},\ }\href@noop {} {\bibfield  {journal} {\bibinfo  {journal}
  {Science}\ }\textbf {\bibinfo {volume} {294}},\ \bibinfo {pages} {1495}
  (\bibinfo {year} {2001})}\BibitemShut {NoStop}%
\bibitem [{\citenamefont {Zhou}\ \emph {et~al.}(2011)\citenamefont {Zhou},
  \citenamefont {Bramwell}, \citenamefont {Cheng}, \citenamefont {Wiebe},
  \citenamefont {Li}, \citenamefont {Balicas}, \citenamefont {Bloxsom},
  \citenamefont {Silverstein}, \citenamefont {Zhou}, \citenamefont {Goodenough}
  \emph {et~al.}}]{zhou2011}%
  \BibitemOpen
  \bibfield  {author} {\bibinfo {author} {\bibfnamefont {H.}~\bibnamefont
  {Zhou}}, \bibinfo {author} {\bibfnamefont {S.}~\bibnamefont {Bramwell}},
  \bibinfo {author} {\bibfnamefont {J.}~\bibnamefont {Cheng}}, \bibinfo
  {author} {\bibfnamefont {C.}~\bibnamefont {Wiebe}}, \bibinfo {author}
  {\bibfnamefont {G.}~\bibnamefont {Li}}, \bibinfo {author} {\bibfnamefont
  {L.}~\bibnamefont {Balicas}}, \bibinfo {author} {\bibfnamefont
  {J.}~\bibnamefont {Bloxsom}}, \bibinfo {author} {\bibfnamefont
  {H.}~\bibnamefont {Silverstein}}, \bibinfo {author} {\bibfnamefont
  {J.}~\bibnamefont {Zhou}}, \bibinfo {author} {\bibfnamefont {J.}~\bibnamefont
  {Goodenough}},  \emph {et~al.},\ }\href@noop {} {\bibfield  {journal}
  {\bibinfo  {journal} {Nature communications}\ }\textbf {\bibinfo {volume}
  {2}},\ \bibinfo {pages} {478} (\bibinfo {year} {2011})}\BibitemShut {NoStop}%
\bibitem [{\citenamefont {Zhou}\ \emph {et~al.}(2012)\citenamefont {Zhou},
  \citenamefont {Cheng}, \citenamefont {Hallas}, \citenamefont {Wiebe},
  \citenamefont {Li}, \citenamefont {Balicas}, \citenamefont {Zhou},
  \citenamefont {Goodenough}, \citenamefont {Gardner},\ and\ \citenamefont
  {Choi}}]{zhou2012chemical}%
  \BibitemOpen
  \bibfield  {author} {\bibinfo {author} {\bibfnamefont {H.}~\bibnamefont
  {Zhou}}, \bibinfo {author} {\bibfnamefont {J.}~\bibnamefont {Cheng}},
  \bibinfo {author} {\bibfnamefont {A.}~\bibnamefont {Hallas}}, \bibinfo
  {author} {\bibfnamefont {C.}~\bibnamefont {Wiebe}}, \bibinfo {author}
  {\bibfnamefont {G.}~\bibnamefont {Li}}, \bibinfo {author} {\bibfnamefont
  {L.}~\bibnamefont {Balicas}}, \bibinfo {author} {\bibfnamefont
  {J.}~\bibnamefont {Zhou}}, \bibinfo {author} {\bibfnamefont {J.}~\bibnamefont
  {Goodenough}}, \bibinfo {author} {\bibfnamefont {J.~S.}\ \bibnamefont
  {Gardner}}, \ and\ \bibinfo {author} {\bibfnamefont {E.}~\bibnamefont
  {Choi}},\ }\href@noop {} {\bibfield  {journal} {\bibinfo  {journal} {Physical
  review letters}\ }\textbf {\bibinfo {volume} {108}},\ \bibinfo {pages}
  {207206} (\bibinfo {year} {2012})}\BibitemShut {NoStop}%
\bibitem [{\citenamefont {Morris}\ \emph {et~al.}(2009)\citenamefont {Morris},
  \citenamefont {Tennant}, \citenamefont {Grigera}, \citenamefont {Klemke},
  \citenamefont {Castelnovo}, \citenamefont {Moessner}, \citenamefont
  {Czternasty}, \citenamefont {Meissner}, \citenamefont {Rule}, \citenamefont
  {Hoffmann}, \citenamefont {Kiefer}, \citenamefont {Gerischer}, \citenamefont
  {Slobinsky},\ and\ \citenamefont {Perry}}]{Morris2009}%
  \BibitemOpen
  \bibfield  {author} {\bibinfo {author} {\bibfnamefont {D.~J.~P.}\
  \bibnamefont {Morris}}, \bibinfo {author} {\bibfnamefont {D.~A.}\
  \bibnamefont {Tennant}}, \bibinfo {author} {\bibfnamefont {S.~A.}\
  \bibnamefont {Grigera}}, \bibinfo {author} {\bibfnamefont {B.}~\bibnamefont
  {Klemke}}, \bibinfo {author} {\bibfnamefont {C.}~\bibnamefont {Castelnovo}},
  \bibinfo {author} {\bibfnamefont {R.}~\bibnamefont {Moessner}}, \bibinfo
  {author} {\bibfnamefont {C.}~\bibnamefont {Czternasty}}, \bibinfo {author}
  {\bibfnamefont {M.}~\bibnamefont {Meissner}}, \bibinfo {author}
  {\bibfnamefont {K.~C.}\ \bibnamefont {Rule}}, \bibinfo {author}
  {\bibfnamefont {J.-U.}\ \bibnamefont {Hoffmann}}, \bibinfo {author}
  {\bibfnamefont {K.}~\bibnamefont {Kiefer}}, \bibinfo {author} {\bibfnamefont
  {S.}~\bibnamefont {Gerischer}}, \bibinfo {author} {\bibfnamefont
  {D.}~\bibnamefont {Slobinsky}}, \ and\ \bibinfo {author} {\bibfnamefont
  {R.~S.}\ \bibnamefont {Perry}},\ }\href {\doibase 10.1126/science.1178868}
  {\bibfield  {journal} {\bibinfo  {journal} {Science}\ }\textbf {\bibinfo
  {volume} {326}},\ \bibinfo {pages} {411} (\bibinfo {year}
  {2009})}\BibitemShut {NoStop}%
\bibitem [{\citenamefont {Jaubert}\ and\ \citenamefont
  {Holdsworth}(2009)}]{jaubert2009nat}%
  \BibitemOpen
  \bibfield  {author} {\bibinfo {author} {\bibfnamefont {L.~D.}\ \bibnamefont
  {Jaubert}}\ and\ \bibinfo {author} {\bibfnamefont {P.~C.}\ \bibnamefont
  {Holdsworth}},\ }\href@noop {} {\bibfield  {journal} {\bibinfo  {journal}
  {Nature Physics}\ }\textbf {\bibinfo {volume} {5}},\ \bibinfo {pages} {258}
  (\bibinfo {year} {2009})}\BibitemShut {NoStop}%
\bibitem [{\citenamefont {Blatter}\ \emph {et~al.}(1994)\citenamefont
  {Blatter}, \citenamefont {Feigel'man}, \citenamefont {Geshkenbein},
  \citenamefont {Larkin},\ and\ \citenamefont {Vinokur}}]{Blatter1994}%
  \BibitemOpen
  \bibfield  {author} {\bibinfo {author} {\bibfnamefont {G.}~\bibnamefont
  {Blatter}}, \bibinfo {author} {\bibfnamefont {M.~V.}\ \bibnamefont
  {Feigel'man}}, \bibinfo {author} {\bibfnamefont {V.~B.}\ \bibnamefont
  {Geshkenbein}}, \bibinfo {author} {\bibfnamefont {A.~I.}\ \bibnamefont
  {Larkin}}, \ and\ \bibinfo {author} {\bibfnamefont {V.~M.}\ \bibnamefont
  {Vinokur}},\ }\href@noop {} {\bibfield  {journal} {\bibinfo  {journal}
  {Reviews of Modern Physics}\ }\textbf {\bibinfo {volume} {66}},\ \bibinfo
  {pages} {1125} (\bibinfo {year} {1994})}\BibitemShut {NoStop}%
\bibitem [{\citenamefont {Sazonov}\ \emph {et~al.}(2012)\citenamefont
  {Sazonov}, \citenamefont {Gukasov}, \citenamefont {Mirebeau},\ and\
  \citenamefont {Bonville}}]{Sazonov2012}%
  \BibitemOpen
  \bibfield  {author} {\bibinfo {author} {\bibfnamefont {A.~P.}\ \bibnamefont
  {Sazonov}}, \bibinfo {author} {\bibfnamefont {A.}~\bibnamefont {Gukasov}},
  \bibinfo {author} {\bibfnamefont {I.}~\bibnamefont {Mirebeau}}, \ and\
  \bibinfo {author} {\bibfnamefont {P.}~\bibnamefont {Bonville}},\ }\href
  {\doibase 10.1103/PhysRevB.85.214420} {\bibfield  {journal} {\bibinfo
  {journal} {Phys. Rev. B}\ }\textbf {\bibinfo {volume} {85}},\ \bibinfo
  {pages} {214420} (\bibinfo {year} {2012})}\BibitemShut {NoStop}%
\bibitem [{\citenamefont {Borzi}\ \emph {et~al.}(2013)\citenamefont {Borzi},
  \citenamefont {Slobinsky},\ and\ \citenamefont {Grigera}}]{Borzi2013}%
  \BibitemOpen
  \bibfield  {author} {\bibinfo {author} {\bibfnamefont {R.~A.}\ \bibnamefont
  {Borzi}}, \bibinfo {author} {\bibfnamefont {D.}~\bibnamefont {Slobinsky}}, \
  and\ \bibinfo {author} {\bibfnamefont {S.~A.}\ \bibnamefont {Grigera}},\
  }\href {\doibase 10.1103/PhysRevLett.111.147204} {\bibfield  {journal}
  {\bibinfo  {journal} {Phys. Rev. Lett.}\ }\textbf {\bibinfo {volume} {111}},\
  \bibinfo {pages} {147204} (\bibinfo {year} {2013})}\BibitemShut {NoStop}%
\bibitem [{\citenamefont {Brooks-Bartlett}\ \emph {et~al.}(2014)\citenamefont
  {Brooks-Bartlett}, \citenamefont {Banks}, \citenamefont {Jaubert},
  \citenamefont {Harman-Clarke},\ and\ \citenamefont
  {Holdsworth}}]{BBartlett2014}%
  \BibitemOpen
  \bibfield  {author} {\bibinfo {author} {\bibfnamefont {M.~E.}\ \bibnamefont
  {Brooks-Bartlett}}, \bibinfo {author} {\bibfnamefont {S.~T.}\ \bibnamefont
  {Banks}}, \bibinfo {author} {\bibfnamefont {L.~D.~C.}\ \bibnamefont
  {Jaubert}}, \bibinfo {author} {\bibfnamefont {A.}~\bibnamefont
  {Harman-Clarke}}, \ and\ \bibinfo {author} {\bibfnamefont {P.~C.~W.}\
  \bibnamefont {Holdsworth}},\ }\href {\doibase 10.1103/PhysRevX.4.011007}
  {\bibfield  {journal} {\bibinfo  {journal} {Phys. Rev. X}\ }\textbf {\bibinfo
  {volume} {4}},\ \bibinfo {pages} {011007} (\bibinfo {year}
  {2014})}\BibitemShut {NoStop}%
\bibitem [{\citenamefont {Guruciaga}\ \emph {et~al.}(2014)\citenamefont
  {Guruciaga}, \citenamefont {Grigera},\ and\ \citenamefont
  {Borzi}}]{Guruciaga2014}%
  \BibitemOpen
  \bibfield  {author} {\bibinfo {author} {\bibfnamefont {P.~C.}\ \bibnamefont
  {Guruciaga}}, \bibinfo {author} {\bibfnamefont {S.~A.}\ \bibnamefont
  {Grigera}}, \ and\ \bibinfo {author} {\bibfnamefont {R.~A.}\ \bibnamefont
  {Borzi}},\ }\href@noop {} {\bibfield  {journal} {\bibinfo  {journal}
  {Physical Review B}\ }\textbf {\bibinfo {volume} {90}},\ \bibinfo {pages}
  {184423} (\bibinfo {year} {2014})}\BibitemShut {NoStop}%
\bibitem [{\citenamefont {Jaubert}\ and\ \citenamefont
  {Moessner}(2015)}]{jaubert2015p}%
  \BibitemOpen
  \bibfield  {author} {\bibinfo {author} {\bibfnamefont {L.~D.~C.}\
  \bibnamefont {Jaubert}}\ and\ \bibinfo {author} {\bibfnamefont
  {R.}~\bibnamefont {Moessner}},\ }\href {\doibase 10.1103/PhysRevB.91.214422}
  {\bibfield  {journal} {\bibinfo  {journal} {Phys. Rev. B}\ }\textbf {\bibinfo
  {volume} {91}},\ \bibinfo {pages} {214422} (\bibinfo {year}
  {2015})}\BibitemShut {NoStop}%
\bibitem [{\citenamefont {{Rau Jeffrey G.}}\ and\ \citenamefont {{Gingras
  Michel J. P.}}(2016)}]{Rau2016}%
  \BibitemOpen
  \bibfield  {author} {\bibinfo {author} {\bibnamefont {{Rau Jeffrey G.}}}\
  and\ \bibinfo {author} {\bibnamefont {{Gingras Michel J. P.}}},\ }\href
  {\doibase http://dx.doi.org/10.1038/ncomms12234 10.1038/ncomms12234}
  {\bibfield  {journal} {\bibinfo  {journal} {Nature Communications}\ }\textbf
  {\bibinfo {volume} {7}},\ \bibinfo {pages} {12234} (\bibinfo {year}
  {2016})}\BibitemShut {NoStop}%
\bibitem [{\citenamefont {Dickman}\ and\ \citenamefont
  {Stell}(1999)}]{Dickman1999}%
  \BibitemOpen
  \bibfield  {author} {\bibinfo {author} {\bibfnamefont {R.}~\bibnamefont
  {Dickman}}\ and\ \bibinfo {author} {\bibfnamefont {G.}~\bibnamefont
  {Stell}},\ }in\ \href@noop {} {\emph {\bibinfo {booktitle} {AIP Conference
  Proceedings}}},\ Vol.\ \bibinfo {volume} {492}\ (\bibinfo {organization}
  {AIP},\ \bibinfo {year} {1999})\ pp.\ \bibinfo {pages} {225--249}\BibitemShut
  {NoStop}%
\bibitem [{\citenamefont {Dickman}(2000)}]{Dickman2000}%
  \BibitemOpen
  \bibfield  {author} {\bibinfo {author} {\bibfnamefont {R.}~\bibnamefont
  {Dickman}},\ }\href@noop {} {\bibfield  {journal} {\bibinfo  {journal}
  {Brazilian Journal of Physics}\ }\textbf {\bibinfo {volume} {30}},\ \bibinfo
  {pages} {711} (\bibinfo {year} {2000})}\BibitemShut {NoStop}%
\bibitem [{\citenamefont {Melko}\ \emph {et~al.}(2001)\citenamefont {Melko},
  \citenamefont {den Hertog},\ and\ \citenamefont {Gingras}}]{Melko2001}%
  \BibitemOpen
  \bibfield  {author} {\bibinfo {author} {\bibfnamefont {R.~G.}\ \bibnamefont
  {Melko}}, \bibinfo {author} {\bibfnamefont {B.~C.}\ \bibnamefont {den
  Hertog}}, \ and\ \bibinfo {author} {\bibfnamefont {M.~J.~P.}\ \bibnamefont
  {Gingras}},\ }\href {\doibase 10.1103/PhysRevLett.87.067203} {\bibfield
  {journal} {\bibinfo  {journal} {Phys. Rev. Lett.}\ }\textbf {\bibinfo
  {volume} {87}},\ \bibinfo {pages} {067203} (\bibinfo {year}
  {2001})}\BibitemShut {NoStop}%
\bibitem [{\citenamefont {Castelnovo}\ \emph {et~al.}(2012)\citenamefont
  {Castelnovo}, \citenamefont {Moessner},\ and\ \citenamefont
  {Sondhi}}]{Castelnovo2012}%
  \BibitemOpen
  \bibfield  {author} {\bibinfo {author} {\bibfnamefont {C.}~\bibnamefont
  {Castelnovo}}, \bibinfo {author} {\bibfnamefont {R.}~\bibnamefont
  {Moessner}}, \ and\ \bibinfo {author} {\bibfnamefont {S.}~\bibnamefont
  {Sondhi}},\ }\href@noop {} {\bibfield  {journal} {\bibinfo  {journal} {Annu.
  Rev. Condens. Matter Phys.}\ }\textbf {\bibinfo {volume} {3}},\ \bibinfo
  {pages} {35} (\bibinfo {year} {2012})}\BibitemShut {NoStop}%
\bibitem [{\citenamefont {Guruciaga}\ \emph {et~al.}(2016)\citenamefont
  {Guruciaga}, \citenamefont {Tarzia}, \citenamefont {Ferreyra}, \citenamefont
  {Cugliandolo}, \citenamefont {Grigera},\ and\ \citenamefont
  {Borzi}}]{Guruciaga2016}%
  \BibitemOpen
  \bibfield  {author} {\bibinfo {author} {\bibfnamefont {P.}~\bibnamefont
  {Guruciaga}}, \bibinfo {author} {\bibfnamefont {M.}~\bibnamefont {Tarzia}},
  \bibinfo {author} {\bibfnamefont {M.}~\bibnamefont {Ferreyra}}, \bibinfo
  {author} {\bibfnamefont {L.}~\bibnamefont {Cugliandolo}}, \bibinfo {author}
  {\bibfnamefont {S.~A.}\ \bibnamefont {Grigera}}, \ and\ \bibinfo {author}
  {\bibfnamefont {R.}~\bibnamefont {Borzi}},\ }\href@noop {} {\bibfield
  {journal} {\bibinfo  {journal} {Physical Review Letters}\ }\textbf {\bibinfo
  {volume} {117}},\ \bibinfo {pages} {167203} (\bibinfo {year}
  {2016})}\BibitemShut {NoStop}%
\bibitem [{\citenamefont {Udagawa}\ \emph {et~al.}(2016)\citenamefont
  {Udagawa}, \citenamefont {Jaubert}, \citenamefont {Castelnovo},\ and\
  \citenamefont {Moessner}}]{Udagawa2016}%
  \BibitemOpen
  \bibfield  {author} {\bibinfo {author} {\bibfnamefont {M.}~\bibnamefont
  {Udagawa}}, \bibinfo {author} {\bibfnamefont {L.}~\bibnamefont {Jaubert}},
  \bibinfo {author} {\bibfnamefont {C.}~\bibnamefont {Castelnovo}}, \ and\
  \bibinfo {author} {\bibfnamefont {R.}~\bibnamefont {Moessner}},\ }\href@noop
  {} {\bibfield  {journal} {\bibinfo  {journal} {Physical Review B}\ }\textbf
  {\bibinfo {volume} {94}},\ \bibinfo {pages} {104416} (\bibinfo {year}
  {2016})}\BibitemShut {NoStop}%
\bibitem [{\citenamefont {Jaubert}(2015)}]{jaubert2015s}%
  \BibitemOpen
  \bibfield  {author} {\bibinfo {author} {\bibfnamefont {L.~D.~C.}\
  \bibnamefont {Jaubert}},\ }\href {\doibase 10.1142/S2010324715400056}
  {\bibfield  {journal} {\bibinfo  {journal} {SPIN}\ }\textbf {\bibinfo
  {volume} {05}},\ \bibinfo {pages} {1540005} (\bibinfo {year}
  {2015})}\BibitemShut {NoStop}%
\bibitem [{\citenamefont {Fennell}\ \emph {et~al.}(2012)\citenamefont
  {Fennell}, \citenamefont {Kenzelmann}, \citenamefont {Roessli}, \citenamefont
  {Haas},\ and\ \citenamefont {Cava}}]{Fennell2012}%
  \BibitemOpen
  \bibfield  {author} {\bibinfo {author} {\bibfnamefont {T.}~\bibnamefont
  {Fennell}}, \bibinfo {author} {\bibfnamefont {M.}~\bibnamefont {Kenzelmann}},
  \bibinfo {author} {\bibfnamefont {B.}~\bibnamefont {Roessli}}, \bibinfo
  {author} {\bibfnamefont {M.~K.}\ \bibnamefont {Haas}}, \ and\ \bibinfo
  {author} {\bibfnamefont {R.~J.}\ \bibnamefont {Cava}},\ }\href {\doibase
  10.1103/PhysRevLett.109.017201} {\bibfield  {journal} {\bibinfo  {journal}
  {Phys. Rev. Lett.}\ }\textbf {\bibinfo {volume} {109}},\ \bibinfo {pages}
  {017201} (\bibinfo {year} {2012})}\BibitemShut {NoStop}%
\bibitem [{\citenamefont {Petit}\ \emph {et~al.}(2012)\citenamefont {Petit},
  \citenamefont {Bonville}, \citenamefont {Robert}, \citenamefont {Decorse},\
  and\ \citenamefont {Mirebeau}}]{Petit2012}%
  \BibitemOpen
  \bibfield  {author} {\bibinfo {author} {\bibfnamefont {S.}~\bibnamefont
  {Petit}}, \bibinfo {author} {\bibfnamefont {P.}~\bibnamefont {Bonville}},
  \bibinfo {author} {\bibfnamefont {J.}~\bibnamefont {Robert}}, \bibinfo
  {author} {\bibfnamefont {C.}~\bibnamefont {Decorse}}, \ and\ \bibinfo
  {author} {\bibfnamefont {I.}~\bibnamefont {Mirebeau}},\ }\href {\doibase
  10.1103/PhysRevB.86.174403} {\bibfield  {journal} {\bibinfo  {journal} {Phys.
  Rev. B}\ }\textbf {\bibinfo {volume} {86}},\ \bibinfo {pages} {174403}
  (\bibinfo {year} {2012})}\BibitemShut {NoStop}%
\bibitem [{\citenamefont {Petit}\ \emph {et~al.}(2016)\citenamefont {Petit},
  \citenamefont {Lhotel}, \citenamefont {Canals}, \citenamefont
  {Ciomaga-Hatnean}, \citenamefont {Ollivier}, \citenamefont {Mutka},
  \citenamefont {Ressouche}, \citenamefont {Wildes}, \citenamefont {Lees},\
  and\ \citenamefont {Balakrishnan}}]{Petit2016}%
  \BibitemOpen
  \bibfield  {author} {\bibinfo {author} {\bibfnamefont {S.}~\bibnamefont
  {Petit}}, \bibinfo {author} {\bibfnamefont {E.}~\bibnamefont {Lhotel}},
  \bibinfo {author} {\bibfnamefont {B.}~\bibnamefont {Canals}}, \bibinfo
  {author} {\bibfnamefont {M.}~\bibnamefont {Ciomaga-Hatnean}}, \bibinfo
  {author} {\bibfnamefont {J.}~\bibnamefont {Ollivier}}, \bibinfo {author}
  {\bibfnamefont {H.}~\bibnamefont {Mutka}}, \bibinfo {author} {\bibfnamefont
  {E.}~\bibnamefont {Ressouche}}, \bibinfo {author} {\bibfnamefont {A.~R.}\
  \bibnamefont {Wildes}}, \bibinfo {author} {\bibfnamefont {M.~R.}\
  \bibnamefont {Lees}}, \ and\ \bibinfo {author} {\bibfnamefont
  {G.}~\bibnamefont {Balakrishnan}},\ }\href@noop {} {\bibfield  {journal}
  {\bibinfo  {journal} {Nature Physics}\ }\textbf {\bibinfo {volume} {12}},\
  \bibinfo {pages} {746} (\bibinfo {year} {2016})}\BibitemShut {NoStop}%
\bibitem [{\citenamefont {Lefran{\c{c}}ois}\ \emph {et~al.}(2017)\citenamefont
  {Lefran{\c{c}}ois}, \citenamefont {Cathelin}, \citenamefont {Lhotel},
  \citenamefont {Robert}, \citenamefont {Lejay}, \citenamefont {Colin},
  \citenamefont {Canals}, \citenamefont {Damay}, \citenamefont {Ollivier},
  \citenamefont {F{\aa}k} \emph {et~al.}}]{Lefrancois2017}%
  \BibitemOpen
  \bibfield  {author} {\bibinfo {author} {\bibfnamefont {E.}~\bibnamefont
  {Lefran{\c{c}}ois}}, \bibinfo {author} {\bibfnamefont {V.}~\bibnamefont
  {Cathelin}}, \bibinfo {author} {\bibfnamefont {E.}~\bibnamefont {Lhotel}},
  \bibinfo {author} {\bibfnamefont {J.}~\bibnamefont {Robert}}, \bibinfo
  {author} {\bibfnamefont {P.}~\bibnamefont {Lejay}}, \bibinfo {author}
  {\bibfnamefont {C.}~\bibnamefont {Colin}}, \bibinfo {author} {\bibfnamefont
  {B.}~\bibnamefont {Canals}}, \bibinfo {author} {\bibfnamefont
  {F.}~\bibnamefont {Damay}}, \bibinfo {author} {\bibfnamefont
  {J.}~\bibnamefont {Ollivier}}, \bibinfo {author} {\bibfnamefont
  {B.}~\bibnamefont {F{\aa}k}},  \emph {et~al.},\ }\href@noop {} {\bibfield
  {journal} {\bibinfo  {journal} {arXiv preprint arXiv:1702.02864}\ } (\bibinfo
  {year} {2017})}\BibitemShut {NoStop}%
\bibitem [{Note1()}]{Note1}%
  \BibitemOpen
  \bibinfo {note} {Though we will not analyze this further, this effect may be
  at the core of the Balance Monopole Liquid Hamiltonian discussed in
  subsequent sections \cite
  {jaubert2015p,jaubert2015s,Khomskii2012}.}\BibitemShut {Stop}%
\bibitem [{\citenamefont {{Khomskii D.I.}}(2012)}]{Khomskii2012}%
  \BibitemOpen
  \bibfield  {author} {\bibinfo {author} {\bibnamefont {{Khomskii D.I.}}},\
  }\href {\doibase http://dx.doi.org/10.1038/ncomms1904 10.1038/ncomms1904}
  {\bibfield  {journal} {\bibinfo  {journal} {Nature Communications}\ }\textbf
  {\bibinfo {volume} {3}},\ \bibinfo {pages} {904} (\bibinfo {year}
  {2012})}\BibitemShut {NoStop}%
\bibitem [{\citenamefont {Fennell}\ \emph {et~al.}(2014)\citenamefont
  {Fennell}, \citenamefont {Kenzelmann}, \citenamefont {Roessli}, \citenamefont
  {Mutka}, \citenamefont {Ollivier}, \citenamefont {Ruminy}, \citenamefont
  {Stuhr}, \citenamefont {Zaharko}, \citenamefont {Bovo}, \citenamefont
  {Cervellino}, \citenamefont {Haas},\ and\ \citenamefont
  {Cava}}]{Fennell2014}%
  \BibitemOpen
  \bibfield  {author} {\bibinfo {author} {\bibfnamefont {T.}~\bibnamefont
  {Fennell}}, \bibinfo {author} {\bibfnamefont {M.}~\bibnamefont {Kenzelmann}},
  \bibinfo {author} {\bibfnamefont {B.}~\bibnamefont {Roessli}}, \bibinfo
  {author} {\bibfnamefont {H.}~\bibnamefont {Mutka}}, \bibinfo {author}
  {\bibfnamefont {J.}~\bibnamefont {Ollivier}}, \bibinfo {author}
  {\bibfnamefont {M.}~\bibnamefont {Ruminy}}, \bibinfo {author} {\bibfnamefont
  {U.}~\bibnamefont {Stuhr}}, \bibinfo {author} {\bibfnamefont
  {O.}~\bibnamefont {Zaharko}}, \bibinfo {author} {\bibfnamefont
  {L.}~\bibnamefont {Bovo}}, \bibinfo {author} {\bibfnamefont {A.}~\bibnamefont
  {Cervellino}}, \bibinfo {author} {\bibfnamefont {M.~K.}\ \bibnamefont
  {Haas}}, \ and\ \bibinfo {author} {\bibfnamefont {R.~J.}\ \bibnamefont
  {Cava}},\ }\href {\doibase 10.1103/PhysRevLett.112.017203} {\bibfield
  {journal} {\bibinfo  {journal} {Phys. Rev. Lett.}\ }\textbf {\bibinfo
  {volume} {112}},\ \bibinfo {pages} {017203} (\bibinfo {year}
  {2014})}\BibitemShut {NoStop}%
\bibitem [{\citenamefont {Borzi}\ \emph {et~al.}(2016)\citenamefont {Borzi},
  \citenamefont {G\'omez~Albarrac\'in}, \citenamefont {Rosales}, \citenamefont
  {Rossini}, \citenamefont {Steppke}, \citenamefont {Prabhakaran},
  \citenamefont {Mackenzie}, \citenamefont {Cabra},\ and\ \citenamefont
  {Grigera}}]{Borzi2016}%
  \BibitemOpen
  \bibfield  {author} {\bibinfo {author} {\bibfnamefont {R.~A.}\ \bibnamefont
  {Borzi}}, \bibinfo {author} {\bibfnamefont {F.~A.}\ \bibnamefont
  {G\'omez~Albarrac\'in}}, \bibinfo {author} {\bibfnamefont {H.~D.}\
  \bibnamefont {Rosales}}, \bibinfo {author} {\bibfnamefont {G.~L.}\
  \bibnamefont {Rossini}}, \bibinfo {author} {\bibfnamefont {A.}~\bibnamefont
  {Steppke}}, \bibinfo {author} {\bibfnamefont {D.}~\bibnamefont
  {Prabhakaran}}, \bibinfo {author} {\bibfnamefont {A.~P.}\ \bibnamefont
  {Mackenzie}}, \bibinfo {author} {\bibfnamefont {D.~C.}\ \bibnamefont
  {Cabra}}, \ and\ \bibinfo {author} {\bibfnamefont {S.~A.}\ \bibnamefont
  {Grigera}},\ }\href {\doibase 10.1038/ncomms12592} {\bibfield  {journal}
  {\bibinfo  {journal} {Nature Communications}\ }\textbf {\bibinfo {volume}
  {7}},\ \bibinfo {pages} {12592} (\bibinfo {year} {2016})}\BibitemShut
  {NoStop}%
\bibitem [{\citenamefont {Penc}\ \emph {et~al.}(2004)\citenamefont {Penc},
  \citenamefont {Shannon},\ and\ \citenamefont {Shiba}}]{Penc2004}%
  \BibitemOpen
  \bibfield  {author} {\bibinfo {author} {\bibfnamefont {K.}~\bibnamefont
  {Penc}}, \bibinfo {author} {\bibfnamefont {N.}~\bibnamefont {Shannon}}, \
  and\ \bibinfo {author} {\bibfnamefont {H.}~\bibnamefont {Shiba}},\ }\href
  {\doibase 10.1103/PhysRevLett.93.197203} {\bibfield  {journal} {\bibinfo
  {journal} {Phys. Rev. Lett.}\ }\textbf {\bibinfo {volume} {93}},\ \bibinfo
  {pages} {197203} (\bibinfo {year} {2004})}\BibitemShut {NoStop}%
\bibitem [{\citenamefont {Bergman}\ \emph {et~al.}(2006)\citenamefont
  {Bergman}, \citenamefont {Shindou}, \citenamefont {Fiete},\ and\
  \citenamefont {Balents}}]{Bergman2006}%
  \BibitemOpen
  \bibfield  {author} {\bibinfo {author} {\bibfnamefont {D.~L.}\ \bibnamefont
  {Bergman}}, \bibinfo {author} {\bibfnamefont {R.}~\bibnamefont {Shindou}},
  \bibinfo {author} {\bibfnamefont {G.~A.}\ \bibnamefont {Fiete}}, \ and\
  \bibinfo {author} {\bibfnamefont {L.}~\bibnamefont {Balents}},\ }\href
  {\doibase 10.1103/PhysRevB.74.134409} {\bibfield  {journal} {\bibinfo
  {journal} {Phys. Rev. B}\ }\textbf {\bibinfo {volume} {74}},\ \bibinfo
  {pages} {134409} (\bibinfo {year} {2006})}\BibitemShut {NoStop}%
\bibitem [{\citenamefont {Tchernyshyov}\ and\ \citenamefont
  {Chern}(2011)}]{tchernyshyov2011lacroix}%
  \BibitemOpen
  \bibfield  {author} {\bibinfo {author} {\bibfnamefont {O.}~\bibnamefont
  {Tchernyshyov}}\ and\ \bibinfo {author} {\bibfnamefont {G.-W.}\ \bibnamefont
  {Chern}}\ }(\bibinfo  {publisher} {Springer},\ \bibinfo {year} {2011})\
  \bibinfo {edition} {2011th}\ ed.,\ Chap.~\bibinfo {chapter} {11}, pp.\
  \bibinfo {pages} {269--291}\BibitemShut {NoStop}%
\bibitem [{\citenamefont {Pomaranski}\ \emph {et~al.}(2013)\citenamefont
  {Pomaranski}, \citenamefont {Yaraskavitch}, \citenamefont {Meng},
  \citenamefont {Ross}, \citenamefont {Noad}, \citenamefont {Dabkowska},
  \citenamefont {Gaulin},\ and\ \citenamefont {Kycia}}]{Pomaranski2013}%
  \BibitemOpen
  \bibfield  {author} {\bibinfo {author} {\bibfnamefont {D.}~\bibnamefont
  {Pomaranski}}, \bibinfo {author} {\bibfnamefont {L.}~\bibnamefont
  {Yaraskavitch}}, \bibinfo {author} {\bibfnamefont {S.}~\bibnamefont {Meng}},
  \bibinfo {author} {\bibfnamefont {K.}~\bibnamefont {Ross}}, \bibinfo {author}
  {\bibfnamefont {H.}~\bibnamefont {Noad}}, \bibinfo {author} {\bibfnamefont
  {H.}~\bibnamefont {Dabkowska}}, \bibinfo {author} {\bibfnamefont
  {B.}~\bibnamefont {Gaulin}}, \ and\ \bibinfo {author} {\bibfnamefont
  {J.}~\bibnamefont {Kycia}},\ }\href@noop {} {\bibfield  {journal} {\bibinfo
  {journal} {Nature Physics}\ }\textbf {\bibinfo {volume} {9}},\ \bibinfo
  {pages} {353} (\bibinfo {year} {2013})}\BibitemShut {NoStop}%
\bibitem [{\citenamefont {Henelius}\ \emph {et~al.}(2016)\citenamefont
  {Henelius}, \citenamefont {Lin}, \citenamefont {Enjalran}, \citenamefont
  {Hao}, \citenamefont {Rau}, \citenamefont {Altosaar}, \citenamefont
  {Flicker}, \citenamefont {Yavors'~kii},\ and\ \citenamefont
  {Gingras}}]{Henelius2016}%
  \BibitemOpen
  \bibfield  {author} {\bibinfo {author} {\bibfnamefont {P.}~\bibnamefont
  {Henelius}}, \bibinfo {author} {\bibfnamefont {T.}~\bibnamefont {Lin}},
  \bibinfo {author} {\bibfnamefont {M.}~\bibnamefont {Enjalran}}, \bibinfo
  {author} {\bibfnamefont {Z.}~\bibnamefont {Hao}}, \bibinfo {author}
  {\bibfnamefont {J.}~\bibnamefont {Rau}}, \bibinfo {author} {\bibfnamefont
  {J.}~\bibnamefont {Altosaar}}, \bibinfo {author} {\bibfnamefont
  {F.}~\bibnamefont {Flicker}}, \bibinfo {author} {\bibfnamefont
  {T.}~\bibnamefont {Yavors'~kii}}, \ and\ \bibinfo {author} {\bibfnamefont
  {M.}~\bibnamefont {Gingras}},\ }\href@noop {} {\bibfield  {journal} {\bibinfo
   {journal} {Physical Review B}\ }\textbf {\bibinfo {volume} {93}},\ \bibinfo
  {pages} {024402} (\bibinfo {year} {2016})}\BibitemShut {NoStop}%
\bibitem [{\citenamefont {Sakakibara}\ \emph {et~al.}(2003)\citenamefont
  {Sakakibara}, \citenamefont {Tayama}, \citenamefont {Hiroi}, \citenamefont
  {Matsuhira},\ and\ \citenamefont {Takagi}}]{Sakakibara2003}%
  \BibitemOpen
  \bibfield  {author} {\bibinfo {author} {\bibfnamefont {T.}~\bibnamefont
  {Sakakibara}}, \bibinfo {author} {\bibfnamefont {T.}~\bibnamefont {Tayama}},
  \bibinfo {author} {\bibfnamefont {Z.}~\bibnamefont {Hiroi}}, \bibinfo
  {author} {\bibfnamefont {K.}~\bibnamefont {Matsuhira}}, \ and\ \bibinfo
  {author} {\bibfnamefont {S.}~\bibnamefont {Takagi}},\ }\href {\doibase
  10.1103/PhysRevLett.90.207205} {\bibfield  {journal} {\bibinfo  {journal}
  {Phys. Rev. Lett.}\ }\textbf {\bibinfo {volume} {90}},\ \bibinfo {pages}
  {207205} (\bibinfo {year} {2003})}\BibitemShut {NoStop}%
\bibitem [{\citenamefont {Barry}\ and\ \citenamefont {Wu}(1989)}]{barry1989}%
  \BibitemOpen
  \bibfield  {author} {\bibinfo {author} {\bibfnamefont {J.}~\bibnamefont
  {Barry}}\ and\ \bibinfo {author} {\bibfnamefont {F.}~\bibnamefont {Wu}},\
  }\href@noop {} {\bibfield  {journal} {\bibinfo  {journal} {International
  Journal of Modern Physics B}\ }\textbf {\bibinfo {volume} {3}},\ \bibinfo
  {pages} {1247} (\bibinfo {year} {1989})}\BibitemShut {NoStop}%
\bibitem [{\citenamefont {Fennell}\ \emph {et~al.}(2009)\citenamefont
  {Fennell}, \citenamefont {Deen}, \citenamefont {Wildes}, \citenamefont
  {Schmalzl}, \citenamefont {Prabhakaran}, \citenamefont {Boothroyd},
  \citenamefont {Aldus}, \citenamefont {McMorrow},\ and\ \citenamefont
  {Bramwell}}]{Fennell2009}%
  \BibitemOpen
  \bibfield  {author} {\bibinfo {author} {\bibfnamefont {T.}~\bibnamefont
  {Fennell}}, \bibinfo {author} {\bibfnamefont {P.~P.}\ \bibnamefont {Deen}},
  \bibinfo {author} {\bibfnamefont {A.~R.}\ \bibnamefont {Wildes}}, \bibinfo
  {author} {\bibfnamefont {K.}~\bibnamefont {Schmalzl}}, \bibinfo {author}
  {\bibfnamefont {D.}~\bibnamefont {Prabhakaran}}, \bibinfo {author}
  {\bibfnamefont {A.~T.}\ \bibnamefont {Boothroyd}}, \bibinfo {author}
  {\bibfnamefont {R.~J.}\ \bibnamefont {Aldus}}, \bibinfo {author}
  {\bibfnamefont {D.~F.}\ \bibnamefont {McMorrow}}, \ and\ \bibinfo {author}
  {\bibfnamefont {S.~T.}\ \bibnamefont {Bramwell}},\ }\href {\doibase
  10.1126/science.1177582} {\bibfield  {journal} {\bibinfo  {journal}
  {Science}\ }\textbf {\bibinfo {volume} {326}},\ \bibinfo {pages} {415}
  (\bibinfo {year} {2009})}\BibitemShut {NoStop}%
\bibitem [{\citenamefont {Chang}\ \emph {et~al.}(2010)\citenamefont {Chang},
  \citenamefont {Su}, \citenamefont {Kao}, \citenamefont {Chou}, \citenamefont
  {Mittal}, \citenamefont {Schneider}, \citenamefont {Br\"uckel}, \citenamefont
  {Balakrishnan},\ and\ \citenamefont {Lees}}]{Chang2010}%
  \BibitemOpen
  \bibfield  {author} {\bibinfo {author} {\bibfnamefont {L.~J.}\ \bibnamefont
  {Chang}}, \bibinfo {author} {\bibfnamefont {Y.}~\bibnamefont {Su}}, \bibinfo
  {author} {\bibfnamefont {Y.-J.}\ \bibnamefont {Kao}}, \bibinfo {author}
  {\bibfnamefont {Y.~Z.}\ \bibnamefont {Chou}}, \bibinfo {author}
  {\bibfnamefont {R.}~\bibnamefont {Mittal}}, \bibinfo {author} {\bibfnamefont
  {H.}~\bibnamefont {Schneider}}, \bibinfo {author} {\bibfnamefont
  {T.}~\bibnamefont {Br\"uckel}}, \bibinfo {author} {\bibfnamefont
  {G.}~\bibnamefont {Balakrishnan}}, \ and\ \bibinfo {author} {\bibfnamefont
  {M.~R.}\ \bibnamefont {Lees}},\ }\href {\doibase 10.1103/PhysRevB.82.172403}
  {\bibfield  {journal} {\bibinfo  {journal} {Phys. Rev. B}\ }\textbf {\bibinfo
  {volume} {82}},\ \bibinfo {pages} {172403} (\bibinfo {year}
  {2010})}\BibitemShut {NoStop}%
\bibitem [{Note2()}]{Note2}%
  \BibitemOpen
  \bibinfo {note} {Along this work we will refer to the analogous of magnetic
  monopoles both as `monopoles' or as `charges'.}\BibitemShut {Stop}%
\bibitem [{Note3()}]{Note3}%
  \BibitemOpen
  \bibinfo {note} {Although $J_{nn} > 0$ in the Ising model, we refer to this
  case as `ferromagnetic', since this is the case when magnetic moments (and
  not pseudospins) are considered \cite {moessner1998rapcomm}.}\BibitemShut
  {Stop}%
\bibitem [{\citenamefont {Henley}(2010)}]{Henley2010}%
  \BibitemOpen
  \bibfield  {author} {\bibinfo {author} {\bibfnamefont {C.~L.}\ \bibnamefont
  {Henley}},\ }\href {\doibase 10.1146/annurev-conmatphys-070909-104138}
  {\bibfield  {journal} {\bibinfo  {journal} {Annual Review of Condensed Matter
  Physics}\ }\textbf {\bibinfo {volume} {1}},\ \bibinfo {pages} {179} (\bibinfo
  {year} {2010})}\BibitemShut {NoStop}%
\bibitem [{\citenamefont {Fennell}\ \emph {et~al.}(2007)\citenamefont
  {Fennell}, \citenamefont {Bramwell}, \citenamefont {McMorrow}, \citenamefont
  {Manuel},\ and\ \citenamefont {Wildes}}]{Fennell2007}%
  \BibitemOpen
  \bibfield  {author} {\bibinfo {author} {\bibfnamefont {T.}~\bibnamefont
  {Fennell}}, \bibinfo {author} {\bibfnamefont {S.}~\bibnamefont {Bramwell}},
  \bibinfo {author} {\bibfnamefont {D.}~\bibnamefont {McMorrow}}, \bibinfo
  {author} {\bibfnamefont {P.}~\bibnamefont {Manuel}}, \ and\ \bibinfo {author}
  {\bibfnamefont {A.}~\bibnamefont {Wildes}},\ }\href@noop {} {\bibfield
  {journal} {\bibinfo  {journal} {Nature Physics}\ }\textbf {\bibinfo {volume}
  {3}},\ \bibinfo {pages} {566} (\bibinfo {year} {2007})}\BibitemShut {NoStop}%
\bibitem [{\citenamefont {den Hertog}\ and\ \citenamefont
  {Gingras}(2000)}]{denHer2000}%
  \BibitemOpen
  \bibfield  {author} {\bibinfo {author} {\bibfnamefont {B.~C.}\ \bibnamefont
  {den Hertog}}\ and\ \bibinfo {author} {\bibfnamefont {M.~J.}\ \bibnamefont
  {Gingras}},\ }\href@noop {} {\bibfield  {journal} {\bibinfo  {journal}
  {Physical review letters}\ }\textbf {\bibinfo {volume} {84}},\ \bibinfo
  {pages} {3430} (\bibinfo {year} {2000})}\BibitemShut {NoStop}%
\bibitem [{\citenamefont {Castelnovo}\ \emph {et~al.}(2011)\citenamefont
  {Castelnovo}, \citenamefont {Moessner},\ and\ \citenamefont
  {Sondhi}}]{Castelnovo2011}%
  \BibitemOpen
  \bibfield  {author} {\bibinfo {author} {\bibfnamefont {C.}~\bibnamefont
  {Castelnovo}}, \bibinfo {author} {\bibfnamefont {R.}~\bibnamefont
  {Moessner}}, \ and\ \bibinfo {author} {\bibfnamefont {S.~L.}\ \bibnamefont
  {Sondhi}},\ }\href {\doibase 10.1103/PhysRevB.84.144435} {\bibfield
  {journal} {\bibinfo  {journal} {Phys. Rev. B}\ }\textbf {\bibinfo {volume}
  {84}},\ \bibinfo {pages} {144435} (\bibinfo {year} {2011})}\BibitemShut
  {NoStop}%
\bibitem [{\citenamefont {Diep}(2004)}]{Diep2004}%
  \BibitemOpen
  \bibfield  {author} {\bibinfo {author} {\bibfnamefont {H.}~\bibnamefont
  {Diep}},\ }\href {https://books.google.com.ar/books?id=eVZmjOvkelUC} {\emph
  {\bibinfo {title} {Frustrated Spin Systems}}}\ (\bibinfo  {publisher} {World
  Scientific},\ \bibinfo {year} {2004})\BibitemShut {NoStop}%
\bibitem [{Note4()}]{Note4}%
  \BibitemOpen
  \bibinfo {note} {Of course, quantum gases do show particle correlations; we
  opted anyway to call our classical system a liquid.}\BibitemShut {Stop}%
\bibitem [{Note5()}]{Note5}%
  \BibitemOpen
  \bibinfo {note} {The charge-charge correlation function was calculated as if
  it were a system of real charges, using the corresponding topological charge
  sitting at the center of a tetrahedron as a variable.}\BibitemShut {Stop}%
\bibitem [{Note6()}]{Note6}%
  \BibitemOpen
  \bibinfo {note} {In order to facilitate its comparison with experiments (see
  next section) we have used for the magnetic moments the form factor
  corresponding to Dy$^{+3}$.}\BibitemShut {Stop}%
\bibitem [{\citenamefont {Bramwell}(2011)}]{bramwell2011lacroix}%
  \BibitemOpen
  \bibfield  {author} {\bibinfo {author} {\bibfnamefont {S.~T.}\ \bibnamefont
  {Bramwell}}\ }(\bibinfo  {publisher} {Springer},\ \bibinfo {year} {2011})\
  \bibinfo {edition} {2011th}\ ed.,\ Chap.~\bibinfo {chapter} {3}, p.~\bibinfo
  {pages} {45}\BibitemShut {NoStop}%
\bibitem [{\citenamefont {Melko}\ and\ \citenamefont
  {Gingras}(2004)}]{Melko2004}%
  \BibitemOpen
  \bibfield  {author} {\bibinfo {author} {\bibfnamefont {R.~G.}\ \bibnamefont
  {Melko}}\ and\ \bibinfo {author} {\bibfnamefont {M.~J.}\ \bibnamefont
  {Gingras}},\ }\href@noop {} {\bibfield  {journal} {\bibinfo  {journal}
  {Journal of Physics: Condensed Matter}\ }\textbf {\bibinfo {volume} {16}},\
  \bibinfo {pages} {R1277} (\bibinfo {year} {2004})}\BibitemShut {NoStop}%
\bibitem [{Note7()}]{Note7}%
  \BibitemOpen
  \bibinfo {note} {Inverting the sign of $J_\square $ leads to a ground state
  populated only by neutral sites and double monopoles in proportion $6:2$;
  this ratio does not change when increasing temperature. We have checked that
  this is, as expected, \protect \emph {another} perfect paramagnet at all
  temperatures.}\BibitemShut {Stop}%
\bibitem [{Note8()}]{Note8}%
  \BibitemOpen
  \bibinfo {note} {D. Slobinsky and R. A. Borzi, in preparation}\BibitemShut
  {NoStop}%
\bibitem [{Note9()}]{Note9}%
  \BibitemOpen
  \bibinfo {note} {A less extreme case, which may include a lower density of
  neutral sites or the combination of neutral and double monopoles, leads to a
  much smaller effect in the same direction.}\BibitemShut {Stop}%
\bibitem [{Note10()}]{Note10}%
  \BibitemOpen
  \bibinfo {note} {We will see later that this could not have happened starting
  from the antiferromagnetic AIAO ground state, since double monopoles do not
  allow for the propagation of dipolar correlations.}\BibitemShut {Stop}%
\bibitem [{\citenamefont {Isakov}\ \emph {et~al.}(2005)\citenamefont {Isakov},
  \citenamefont {Moessner},\ and\ \citenamefont {Sondhi}}]{Isakov2005}%
  \BibitemOpen
  \bibfield  {author} {\bibinfo {author} {\bibfnamefont {S.~V.}\ \bibnamefont
  {Isakov}}, \bibinfo {author} {\bibfnamefont {R.}~\bibnamefont {Moessner}}, \
  and\ \bibinfo {author} {\bibfnamefont {S.~L.}\ \bibnamefont {Sondhi}},\
  }\href {\doibase 10.1103/PhysRevLett.95.217201} {\bibfield  {journal}
  {\bibinfo  {journal} {Phys. Rev. Lett.}\ }\textbf {\bibinfo {volume} {95}},\
  \bibinfo {pages} {217201} (\bibinfo {year} {2005})}\BibitemShut {NoStop}%
\bibitem [{\citenamefont {Sen}\ \emph {et~al.}(2013)\citenamefont {Sen},
  \citenamefont {Moessner},\ and\ \citenamefont {Sondhi}}]{Sen2013}%
  \BibitemOpen
  \bibfield  {author} {\bibinfo {author} {\bibfnamefont {A.}~\bibnamefont
  {Sen}}, \bibinfo {author} {\bibfnamefont {R.}~\bibnamefont {Moessner}}, \
  and\ \bibinfo {author} {\bibfnamefont {S.~L.}\ \bibnamefont {Sondhi}},\
  }\href {\doibase 10.1103/PhysRevLett.110.107202} {\bibfield  {journal}
  {\bibinfo  {journal} {Phys. Rev. Lett.}\ }\textbf {\bibinfo {volume} {110}},\
  \bibinfo {pages} {107202} (\bibinfo {year} {2013})}\BibitemShut {NoStop}%
\bibitem [{\citenamefont {Bramwell}(2017)}]{bramwell2017harmonic}%
  \BibitemOpen
  \bibfield  {author} {\bibinfo {author} {\bibfnamefont {S.~T.}\ \bibnamefont
  {Bramwell}},\ }\href@noop {} {\bibfield  {journal} {\bibinfo  {journal}
  {Nature communications}\ }\textbf {\bibinfo {volume} {8}},\ \bibinfo {pages}
  {2088} (\bibinfo {year} {2017})}\BibitemShut {NoStop}%
\bibitem [{\citenamefont {Villain}\ \emph {et~al.}(1980)\citenamefont
  {Villain}, \citenamefont {Bidaux}, \citenamefont {Carton},\ and\
  \citenamefont {Conte}}]{villain1980}%
  \BibitemOpen
  \bibfield  {author} {\bibinfo {author} {\bibfnamefont {J.}~\bibnamefont
  {Villain}}, \bibinfo {author} {\bibfnamefont {R.}~\bibnamefont {Bidaux}},
  \bibinfo {author} {\bibfnamefont {J.-P.}\ \bibnamefont {Carton}}, \ and\
  \bibinfo {author} {\bibfnamefont {R.}~\bibnamefont {Conte}},\ }\href@noop {}
  {\bibfield  {journal} {\bibinfo  {journal} {Journal de Physique}\ }\textbf
  {\bibinfo {volume} {41}},\ \bibinfo {pages} {1263} (\bibinfo {year}
  {1980})}\BibitemShut {NoStop}%
\bibitem [{\citenamefont {Lacroix}\ \emph {et~al.}(2011)\citenamefont
  {Lacroix}, \citenamefont {Mendels},\ and\ \citenamefont
  {Mila}}]{Lacroix2011}%
  \BibitemOpen
  \bibinfo {editor} {\bibfnamefont {C.}~\bibnamefont {Lacroix}}, \bibinfo
  {editor} {\bibfnamefont {P.}~\bibnamefont {Mendels}}, \ and\ \bibinfo
  {editor} {\bibfnamefont {F.}~\bibnamefont {Mila}},\ eds.,\ \href {\doibase
  10.1007/978-3-642-10589-0} {\emph {\bibinfo {title} {{Introduction to
  Frustrated Magnetism: Materials, Experiments, Theory (Springer Series in
  Solid-State Sciences)}}}},\ \bibinfo {edition} {2011th}\ ed.\ (\bibinfo
  {publisher} {Springer},\ \bibinfo {year} {2011})\BibitemShut {NoStop}%
\bibitem [{\citenamefont {Machida}\ \emph {et~al.}(2007)\citenamefont
  {Machida}, \citenamefont {Nakatsuji}, \citenamefont {Maeno}, \citenamefont
  {Tayama}, \citenamefont {Sakakibara},\ and\ \citenamefont
  {Onoda}}]{machida2007unconventional}%
  \BibitemOpen
  \bibfield  {author} {\bibinfo {author} {\bibfnamefont {Y.}~\bibnamefont
  {Machida}}, \bibinfo {author} {\bibfnamefont {S.}~\bibnamefont {Nakatsuji}},
  \bibinfo {author} {\bibfnamefont {Y.}~\bibnamefont {Maeno}}, \bibinfo
  {author} {\bibfnamefont {T.}~\bibnamefont {Tayama}}, \bibinfo {author}
  {\bibfnamefont {T.}~\bibnamefont {Sakakibara}}, \ and\ \bibinfo {author}
  {\bibfnamefont {S.}~\bibnamefont {Onoda}},\ }\href@noop {} {\bibfield
  {journal} {\bibinfo  {journal} {Physical review letters}\ }\textbf {\bibinfo
  {volume} {98}},\ \bibinfo {pages} {057203} (\bibinfo {year}
  {2007})}\BibitemShut {NoStop}%
\bibitem [{\citenamefont {Slobinsky}\ \emph {et~al.}(2018)\citenamefont
  {Slobinsky}, \citenamefont {Baglietto},\ and\ \citenamefont
  {Borzi}}]{Slobinskyinprep}%
  \BibitemOpen
  \bibfield  {author} {\bibinfo {author} {\bibfnamefont {D.}~\bibnamefont
  {Slobinsky}}, \bibinfo {author} {\bibfnamefont {G.}~\bibnamefont
  {Baglietto}}, \ and\ \bibinfo {author} {\bibfnamefont {R.~A.}\ \bibnamefont
  {Borzi}},\ }\href@noop {} {\bibfield  {journal} {\bibinfo  {journal} {in
  preparation}\ } (\bibinfo {year} {2018})}\BibitemShut {NoStop}%
\bibitem [{\citenamefont {Xie}\ \emph {et~al.}(2015)\citenamefont {Xie},
  \citenamefont {Du}, \citenamefont {Yan},\ and\ \citenamefont
  {Liu}}]{Xie2015}%
  \BibitemOpen
  \bibfield  {author} {\bibinfo {author} {\bibfnamefont {Y.-L.}\ \bibnamefont
  {Xie}}, \bibinfo {author} {\bibfnamefont {Z.-Z.}\ \bibnamefont {Du}},
  \bibinfo {author} {\bibfnamefont {Z.-B.}\ \bibnamefont {Yan}}, \ and\
  \bibinfo {author} {\bibfnamefont {J.-M.}\ \bibnamefont {Liu}},\ }\href@noop
  {} {\bibfield  {journal} {\bibinfo  {journal} {Scientific reports}\ }\textbf
  {\bibinfo {volume} {5}} (\bibinfo {year} {2015})}\BibitemShut {NoStop}%
\bibitem [{\citenamefont {Baez}\ and\ \citenamefont {Borzi}(2017)}]{Baez16}%
  \BibitemOpen
  \bibfield  {author} {\bibinfo {author} {\bibfnamefont {M.~L.}\ \bibnamefont
  {Baez}}\ and\ \bibinfo {author} {\bibfnamefont {R.~A.}\ \bibnamefont
  {Borzi}},\ }\href {http://stacks.iop.org/0953-8984/29/i=5/a=055806}
  {\bibfield  {journal} {\bibinfo  {journal} {Journal of Physics: Condensed
  Matter}\ }\textbf {\bibinfo {volume} {29}},\ \bibinfo {pages} {055806}
  (\bibinfo {year} {2017})}\BibitemShut {NoStop}%
\bibitem [{Note11()}]{Note11}%
  \BibitemOpen
  \bibinfo {note} {\protect \url
  {http://lamp.tu-graz.ac.at/~hadley/ss1/crystaldiffraction/atomicformfactors/formfactors.php}}\BibitemShut
  {NoStop}%
\bibitem [{\citenamefont {Moessner}(1998)}]{moessner1998rapcomm}%
  \BibitemOpen
  \bibfield  {author} {\bibinfo {author} {\bibfnamefont {R.}~\bibnamefont
  {Moessner}},\ }\href@noop {} {\bibfield  {journal} {\bibinfo  {journal}
  {Physical Review B}\ }\textbf {\bibinfo {volume} {57}},\ \bibinfo {pages}
  {R5587} (\bibinfo {year} {1998})}\BibitemShut {NoStop}%
\end{thebibliography}%

\end{document}